\documentclass[ALICE,manyauthors]{cernphprep}
\usepackage[comma,square,numbers,sort&compress]{natbib}
\usepackage{hyperref}
\usepackage{lineno}
\usepackage{xspace}
\usepackage{graphicx}
\usepackage{xcolor}
\usepackage[T1]{fontenc}
\usepackage{orcidlink}

\begin{document}
%

\newcommand{\pp}           {pp\xspace}
\newcommand{\ppbar}        {\mbox{$\mathrm {p\overline{p}}$}\xspace}
\newcommand{\XeXe}         {\mbox{Xe--Xe}\xspace}
\newcommand{\PbPb}         {\mbox{Pb--Pb}\xspace}
\newcommand{\pA}           {\mbox{pA}\xspace}
\newcommand{\pPb}          {\mbox{p--Pb}\xspace}
\newcommand{\AuAu}         {\mbox{Au--Au}\xspace}
\newcommand{\dAu}          {\mbox{d--Au}\xspace}

\newcommand{\s}            {\ensuremath{\sqrt{s}}\xspace}
\newcommand{\snn}          {\ensuremath{\sqrt{s_{\mathrm{NN}}}}\xspace}
\newcommand{\pt}           {\ensuremath{p_{\rm T}}\xspace}
\newcommand{\meanpt}       {$\langle p_{\mathrm{T}}\rangle$\xspace}
\newcommand{\ycms}         {\ensuremath{y_{\rm CMS}}\xspace}
\newcommand{\ylab}         {\ensuremath{y_{\rm lab}}\xspace}
\newcommand{\etarange}[1]  {\mbox{$\left | \eta \right |~<~#1$}}
\newcommand{\yrange}[1]    {\mbox{$\left | y \right |~<~#1$}}
\newcommand{\dndy}         {\ensuremath{\mathrm{d}N_\mathrm{ch}/\mathrm{d}y}\xspace}
\newcommand{\dndeta}       {\ensuremath{\mathrm{d}N_\mathrm{ch}/\mathrm{d}\eta}\xspace}
\newcommand{\avdndeta}     {\ensuremath{\langle\dndeta\rangle}\xspace}
\newcommand{\dNdy}         {\ensuremath{\mathrm{d}N_\mathrm{ch}/\mathrm{d}y}\xspace}
\newcommand{\Npart}        {\ensuremath{N_\mathrm{part}}\xspace}
\newcommand{\Ncoll}        {\ensuremath{N_\mathrm{coll}}\xspace}
\newcommand{\dEdx}         {\ensuremath{\textrm{d}E/\textrm{d}x}\xspace}
\newcommand{\RpPb}         {\ensuremath{R_{\rm pPb}}\xspace}

\newcommand{\nineH}        {$\sqrt{s}~=~0.9$~Te\kern-.1emV\xspace}
\newcommand{\seven}        {$\sqrt{s}~=~7$~Te\kern-.1emV\xspace}
\newcommand{\twoH}         {$\sqrt{s}~=~0.2$~Te\kern-.1emV\xspace}
\newcommand{\twosevensix}  {$\sqrt{s}~=~2.76$~Te\kern-.1emV\xspace}
\newcommand{\five}         {$\sqrt{s}~=~5.02$~Te\kern-.1emV\xspace}
\newcommand{\twosevensixnn}{$\sqrt{s_{\mathrm{NN}}}~=~2.76$~Te\kern-.1emV\xspace}
\newcommand{\fivenn}       {$\sqrt{s_{\mathrm{NN}}}~=~5.02$~Te\kern-.1emV\xspace}
\newcommand{\LT}           {L{\'e}vy-Tsallis\xspace}
\newcommand{\GeVc}         {$\mathrm{GeV}/c$}
\newcommand{\MeVc}         {Me\kern-.1emV/$c$\xspace}
\newcommand{\TeV}          {Te\kern-.1emV\xspace}
\newcommand{\GeV}          {Ge\kern-.1emV\xspace}
\newcommand{\MeV}          {Me\kern-.1emV\xspace}
\newcommand{\GeVmass}      {$\mathrm{GeV}/c^2$}
\newcommand{\MeVmass}      {Me\kern-.1emV/$c^2$\xspace}
\newcommand{\lumi}         {\ensuremath{\mathcal{L}}\xspace}

\newcommand{\ITS}          {\rm{ITS}\xspace}
\newcommand{\TOF}          {\rm{TOF}\xspace}
\newcommand{\ZDC}          {\rm{ZDC}\xspace}
\newcommand{\ZDCs}         {\rm{ZDCs}\xspace}
\newcommand{\ZNA}          {\rm{ZNA}\xspace}
\newcommand{\ZNC}          {\rm{ZNC}\xspace}
\newcommand{\SPD}          {\rm{SPD}\xspace}
\newcommand{\SDD}          {\rm{SDD}\xspace}
\newcommand{\SSD}          {\rm{SSD}\xspace}
\newcommand{\TPC}          {\rm{TPC}\xspace}
\newcommand{\TRD}          {\rm{TRD}\xspace}
\newcommand{\VZERO}        {\rm{V0}\xspace}
\newcommand{\VZEROA}       {\rm{V0A}\xspace}
\newcommand{\VZEROC}       {\rm{V0C}\xspace}
\newcommand{\TZERO}        {\rm{T0}\xspace}
\newcommand{\TZEROA}       {\rm{T0A}\xspace}
\newcommand{\TZEROC}       {\rm{T0C}\xspace}
\newcommand{\Vdecay} 	   {\ensuremath{{\rm V}^{0}}\xspace}

\newcommand{\ee}           {\ensuremath{e^{+}e^{-}}} 
\newcommand{\pip}          {\ensuremath{\pi^{+}}\xspace}
\newcommand{\pim}          {\ensuremath{\pi^{-}}\xspace}
\newcommand{\pipm}          {\ensuremath{\pi^{\pm}}\xspace}
\newcommand{\kap}          {\ensuremath{\rm{K}^{+}}\xspace}
\newcommand{\kam}          {\ensuremath{\rm{K}^{-}}\xspace}
\newcommand{\kapm}          {\ensuremath{\rm{K}^{\pm}}\xspace}
\newcommand{\pbar}         {\ensuremath{\rm\overline{p}}\xspace}
\newcommand{\kzero}        {\ensuremath{{\rm K}^{0}_{\rm{S}}}\xspace}
\newcommand{\lmb}          {\ensuremath{\Lambda}\xspace}
\newcommand{\almb}         {\ensuremath{\overline{\Lambda}}\xspace}
\newcommand{\Om}           {\ensuremath{\Omega^-}\xspace}
\newcommand{\Mo}           {\ensuremath{\overline{\Omega}^+}\xspace}
\newcommand{\X}            {\ensuremath{\Xi^-}\xspace}
\newcommand{\Ix}           {\ensuremath{\overline{\Xi}^+}\xspace}
\newcommand{\Xis}          {\ensuremath{\Xi^{\pm}}\xspace}
\newcommand{\Oms}          {\ensuremath{\Omega^{\pm}}\xspace}
\newcommand{\degree}       {\ensuremath{^{\rm o}}\xspace}

\newcommand{\uQpp}{$\langle{M}\rangle^{\textrm{pp}}\langle{\langle{\vec{u}_{n}\cdot\frac{\vec{Q}^{*}_{n}}{M}}\rangle}\rangle^{\rm{pp}}$}
\newcommand{\uQAA}{$\langle{M}\rangle^{\textrm{AA}}\langle{\langle{\vec{u}_{n}\cdot\frac{\vec{Q}^{*}_{n}}{M}}\rangle}\rangle^{\rm{AA}}$}

\newcommand{\vn}[1]{$v_{#1}$}
\newcommand{\vnTwo}[1]{$v_{#1}\{2\}$}
\newcommand{\vnTwoGap}[2]{$v_{#1}\{2,|\Delta\eta|>#2\}$}
\newcommand{\vnFour}[1]{$v_{#1}\{4\}$}
\newcommand{\vnFourGap}[2]{$v_{#1}\{4,|\Delta\eta|>#2\}$}

\newcommand{\cn}[1]{$c_{#1}$}
\newcommand{\cnTwo}[1]{$c_{#1}\{2\}$}
\newcommand{\cnFour}[1]{$c_{#1}\{4\}$}

\newcommand{\dn}[1]{$d_{#1}$}
\newcommand{\dnTwo}[1]{$d_{#1}\{2\}$}
\newcommand{\dnFour}[1]{$d_{#1}\{4\}$}

\newcommand{\mvn}[1]{v_{#1}}
\newcommand{\mvnTwo}[1]{v_{#1}\{2\}}
\newcommand{\mvnFour}[1]{v_{#1}\{4\}}

\newcommand{\mcn}[1]{c_{#1}}
\newcommand{\mcnTwo}[1]{c_{#1}\{2\}}
\newcommand{\mcnFour}[1]{c_{#1}\{4\}}

\newcommand{\mdn}[1]{d_{#1}}
\newcommand{\mdnTwo}[1]{d_{#1}\{2\}}
\newcommand{\mdnFour}[1]{d_{#1}\{4\}}

\newcommand{\cor}[2]{\langle {#1} \rangle_{#2}}
\newcommand{\avcor}[2]{\langle\langle {#1} \rangle\rangle_{#2}}

\newcommand{\rawvtwo}{$v_{2}$}
\newcommand{\rawvthree}{$v_{3}$}
\newcommand{\rawvfour}{$v_{4}$}
\newcommand{\rawvfive}{$v_{5}$}
\newcommand{\rawvn}{$v_{n}$}

\newcommand{\rawvtwonq}{$v_{2}/n_{q}$}
\newcommand{\rawvthreenq}{$v_{3}/n_{q}$}
\newcommand{\rawvfournq}{$v_{4}/n_{q}$}
\newcommand{\rawvfivenq}{$v_{5}/n_{q}$}
\newcommand{\rawvnnq}{$v_{n}/n_{q}$}

\newcommand{\rawvtwotot}{$v_{2}^\mathrm{tot}$}
\newcommand{\rawvtwosig}{$v_{2}^\mathrm{sig}$}
\newcommand{\rawvtwobg}{$v_{2}^\mathrm{bg}$}

\newcommand{\vtwo}{$v_{2}^{\mathrm{sub}}$}
\newcommand{\vthree}{$v_{3}^{\mathrm{sub}}$}
\newcommand{\vfour}{$v_{4}^{\mathrm{sub}}$}
\newcommand{\vfive}{$v_{5}^{\mathrm{sub}}$}

\newcommand{\vtwonq}{$v_{2}^{\mathrm{sub}}/n_{q}$}
\newcommand{\vthreenq}{$v_{3}^{\mathrm{sub}}/n_{q}$}
\newcommand{\vfournq}{$v_{4}^{\mathrm{sub}}/n_{q}$}
\newcommand{\vfivenq}{$v_{5}^{\mathrm{sub}}/n_{q}$}
\newcommand{\vnnq}{$v_{n}^{\mathrm{sub}}/n_{q}$}

\newcommand{\etagap}{$|\Delta\eta|>0$}

\newcommand{\pT}{$p_{\mathrm{T}}$}
\newcommand{\mpT}{p_{\mathrm{T}}}
\newcommand{\minv}{$m_{\mathrm{inv}}$}
\newcommand{\mminv}{m_{\mathrm{inv}}}
\newcommand{\nq}{$n_\mathrm{q}$}
\newcommand{\pTnq}{$p_{\mathrm{T}}/n_\mathrm{q}$}

\newcommand{\h}{~$\mathrm{h}^{\pm}$}
\newcommand{\e}{~$\mathrm{e}^{\pm}$}
\newcommand{\muon}{~$\mathrm{\mu}^{\pm}$}
\newcommand{\pion}{~$\mathrm{\pi}^{\pm}$}
\newcommand{\kaon}{~$\mathrm{K}^{\pm}$}
\newcommand{\proton}{~$\mathrm{p}(\bar{\mathrm{p}})$}
\newcommand{\phiMeson}{~$\mathrm{\phi}$}
\newcommand{\Kos}{~$\mathrm{K_{S}^{0}}$}
\newcommand{\lambdas}{~$\Lambda(\bar{\Lambda})$}
\newcommand{\lambdasZ}{$\Lambda(\bar{\Lambda})$}
\newcommand{\vo}{~$\mathrm{V^{0}}$}
\newcommand{\voZ}{$\mathrm{V^{0}}$}
\newcommand{\omegaM}{~$\mathrm{\Omega^{\mp}}$}
\newcommand{\xiM}{~$\mathrm{\Xi^{\mp}}$}

\newcommand{\todo}[1]{\textcolor{purple}{[TODO: \emph{#1}]}}
\newcommand{\done}[1]{\textcolor{blue}{[DONE: \emph{#1}]}}
\newcommand{\new}[1]{\textcolor{red}{#1}}
\newcommand{\old}[1]{\textcolor{orange}{#1}}
\newcommand{\change}[1]{\textcolor{olive}{[CHANGE: #1]}}
\newcommand{\ask}[1]{\textcolor{magenta}{\bf [ #1]}}

\begin{titlepage}
\PHyear{2022}       
\PHnumber{122}      
\PHdate{07 June}  

\title{Anisotropic flow and flow fluctuations of identified hadrons in Pb--Pb collisions at $\mathbf{\sqrt{s_{\mathrm{NN}}} = 5.02}$~TeV}
\ShortTitle{Flow and flow fluctuations for different particle species at the LHC}   

\Collaboration{ALICE Collaboration\thanks{See Appendix~\ref{app:collab} for the list of collaboration members}}
\ShortAuthor{ALICE Collaboration} 

\begin{abstract}

The first measurements of elliptic flow of \pipm{}, \kapm{}, p+\pbar{}, \kzero{}, \lmb{}+\almb{}, $\phi$, \X{}+\Ix{}, and \Om{}+\Mo{} using multiparticle cumulants in Pb--Pb collisions at $\sqrt{s_{\rm NN}} = 5.02$~TeV are presented. Results obtained with two- ($v_2\{2\}$) and four-particle cumulants ($v_2\{4\}$) are shown as a function of transverse momentum, \pt{}, for various collision centrality intervals. Combining the data for both $v_2\{2\}$ and $v_2\{4\}$ also allows us to report the first measurements of the mean elliptic flow, elliptic flow fluctuations, and relative elliptic flow fluctuations for various hadron species. These observables probe the event-by-event eccentricity fluctuations in the initial state and the contributions from the dynamic evolution of the expanding quark--gluon plasma. The characteristic features observed in previous \pt-differential anisotropic flow measurements for identified hadrons with two-particle correlations, namely the mass ordering at low \pt{} and the approximate scaling with the number of constituent quarks at intermediate \pt{}, are similarly present in the four-particle correlations and the combinations of $v_2\{2\}$ and $v_2\{4\}$. In addition, a particle species dependence of flow fluctuations is observed that could indicate a significant contribution from final state hadronic interactions. The comparison between experimental measurements and 
CoLBT model calculations, which combine the various physics processes of hydrodynamics, quark coalescence, and jet fragmentation, illustrates their importance over a wide \pt{} range. 

\end{abstract}
\end{titlepage}

\setcounter{page}{2} 



\section{Introduction}
\label{Sec:Intro}

The primary goal of the ultra-relativistic heavy-ion collision programme at the Large Hadron Collider (LHC) is to study the properties of the quark--gluon plasma (QGP), a novel state of strongly interacting matter at high temperatures and energy densities~\cite{Shuryak:1978ij,Shuryak:1980tp}. Studies of the azimuthal anisotropy of particle production have contributed significantly to the characterization of the system created in heavy-ion collisions~\cite{Ollitrault:1992bk,Voloshin:2008dg}.
Anisotropic flow reflects the conversion of the initial state spatial anisotropy into final state anisotropies in momentum space. This translation is facilitated by interactions between the constituents of the quark--gluon plasma (QGP)~\cite{Shuryak:1984nq,Cleymans:1985wb,Bass:1998vz} and at later stages, after hadronisation, between the produced particles. Anisotropic flow is quantified by studying the azimuthal distribution of particles emitted in the plane transverse to the beam direction~\cite{Ollitrault:1992bk}. This is usually expressed in terms of a Fourier series in the azimuthal angle $\varphi$~\cite{Voloshin:1994mz,Poskanzer:1998yz} according to
\begin{equation}
E\frac{\mathrm{d}^3N}{\mathrm{d}p^3} = \frac{1}{2\pi}\frac{\mathrm{d}^2N}{p_{\mathrm{T}}\mathrm{d}p_{\mathrm{T}}\mathrm{d}\eta} \Big\{1 + 2\sum_{\rm n=1}^{\infty} v_{\rm n}(p_{\mathrm{T}},\eta) \cos[{\rm n}(\varphi - \Psi_{\rm n})]\Big\},
\label{Eq:Fourier}
\end{equation}
\noindent where $E$, $N$, $p$, \pt{}, $\varphi$, and $\eta$ are the energy, yield, momentum, transverse momentum, azimuthal angle, and pseudorapidity of particles, respectively, and $\Psi_{\rm n}$ is the azimuthal angle of the symmetry plane of order n~\cite{Alver:2010gr,Alver:2010dn}. The $v_{\rm n}$ coefficients are given by 
\begin{equation}
v_{\rm n} = \langle{\cos[{\rm n}(\varphi - \Psi_{\rm n})]}\rangle,
\label{Eq:vn}
\end{equation}
where $\langle\rangle$ denote an average over all particles in a single event. The second Fourier coefficient, $v_2$, is usually referred to as elliptic flow. It is the dominant harmonic in heavy-ion collisions with large values of impact parameter (i.e. non-central collisions). Its value is sensitive to some of the basic transport coefficients of the QGP, e.g. the shear viscosity over entropy density ratio ($\eta/s$). The study of elliptic flow has been instrumental in establishing the strongly-coupled QGP paradigm, first in collisions at the Relativistic Heavy Ion Collider (RHIC)~\cite{Arsene:2004fa,Adcox:2004mh,Back:2004je,Adams:2005dq} and, since 2010, in collisions at the Large Hadron Collider (LHC)~\cite{ALICE:2011ab,ATLAS:2012at,Chatrchyan:2013kba}.

Around the start of the LHC heavy-ion program, it was realised that elliptic flow is also a sensitive probe of the initial state of heavy-ion collisions~\cite{Gardim:2011xv}. Its magnitude fluctuates from one event to the other, reflecting the event-by-event fluctuating energy-density profiles of the nuclear overlap region prior to the formation of the QGP. Initial event-by-event geometry fluctuations lead to the fluctuations of even-harmonic anisotropic flow and generate non-zero odd harmonics. In fact, the initial geometry fluctuations lead to $\langle v_{\rm n}^{\rm k} \rangle \neq \langle v_{\rm n}\rangle^{\rm k}$ and the development of different order symmetry planes $\Psi_{\rm n}$ in different kinematic regions in \pt{} or $\eta$~\cite{CMS:2015xmx,ALICE:2017lyf,ATLAS:2017rij}. Thus, a comprehensive investigation of the final state flow fluctuations is crucial for understanding the event-by-event initial geometry fluctuations and their impact on the system dynamic evolution.
Studies of charged particles in Pb--Pb collisions at LHC energies indicated non-Gaussian initial state fluctuations and, consequently, made it possible to constrain their probability distribution function (p.d.f.)~\cite{Acharya:2018lmh,Sirunyan:2017fts}. 
Studies of flow fluctuations have so far been performed both experimentally and in theoretical model calculations for the measurements integrated over a large kinematic range~\cite{ALICE:2012vgf,Abelev:2014pua,ALICE:2016cti,ALICE:2018yph}. On the other hand, a \pt-differential study, and in particular with identified hadrons, has not been done before. These studies can provide insights on the interplay between the expansion of the system and its late-stage, highly-dissipative hadronic phase, as well as particle production mechanisms. 

Similar studies in the past for various flow coefficients have been pivotal in establishing the need to include viscous corrections in hydrodynamic models and, consequently, in constraining the value of $\eta/s$ to be very close to the conjectured lower limit of $1/4\pi$ calculated for infinitely strongly coupled gauge theories via the AdS/CFT correspondence~\cite{Kovtun:2004de}. Detailed studies of how anisotropic flow develops for different particle species as a function of \pt{} for various centrality intervals (i.e. an estimate of the degree of overlap between the two colliding nuclei) of Pb--Pb collisions at LHC energies~\cite{Abelev:2014pua,ALICE:2016cti,ALICE:2018yph} confirmed a number of qualitative features already observed at RHIC~\cite{Arsene:2004fa,Adcox:2004mh,Back:2004je,Adams:2005dq}: the mass ordering of $v_{\rm n}$ at low \pt{} and the particle type (i.e. mesons versus baryons) grouping at intermediate \pt{}. The former originates from the interplay between radial flow and the anisotropic expansion of the system because of a thermalized expanding source with a common flow velocity for the produced particles~\cite{Huovinen:2001cy}, while the latter is interpreted as an indication of hadron formation via quark coalescence in this momentum range~\cite{Voloshin:2002wa,Molnar:2003ff}. These studies also revealed, for the first time, similar qualitative features in the 1\% most central Pb--Pb collisions~\cite{ALICE:2016cti}, a category of events known as ultracentral with no prevailing ellipsoidal geometry. In addition, new results on the non-linear flow modes of higher harmonics~\cite{ALICE:2019xkq} illustrated unambiguously that the aforementioned features can still be observed after the non-linear response of the system, the latter being proportional to the product of lower-order initial spatial anisotropies~\cite{Bhalerao:2014xra,Yan:2015jma,ALICE:2017fcd,ALICE:2020sup}.

All these studies relied on measuring various flow harmonics using variations of two-particle correlation techniques. One of the disadvantages of such approaches is that they are sensitive to non-flow effects, i.e. correlations between (mainly two) particles not associated with the common symmetry plane. In order to suppress such contributions, one could measure multiparticle cumulants which need a larger data sample to reach the same level of uncertainties as to their two-particle counterpart measurements. This is possible now by combining the entire data set of Pb--Pb collisions at the centre-of-mass energy per nucleon pair $\sqrt{s_{\mathrm{NN}}} = 5.02$~TeV from the LHC Run 2. Furthermore, the usage of higher-order cumulants opens up the possibility to study, for the first time, the particle species dependence of flow fluctuations. 

The first measurements of elliptic flow and flow fluctuations using two- and four-particle cumulants for \pipm{}, \kapm{}, p+\pbar{}, \kzero{}, \lmb{}+\almb{}, $\phi$, \X{}+\Ix{}, and \Om{}+\Mo{} in Pb--Pb collisions at $\sqrt{s_{\mathrm{NN}}} = 5.02$~TeV are presented in this article. Results obtained with the generic framework~\cite{Bilandzic:2013kga,Huo:2017nms,Moravcova:2020wnf}, which corrects detector inefficiencies and non-uniformities in the azimuthal acceptance, are reported for a wide range of transverse momenta ($0.2<\pt<6$~\GeVc{}) in the 10--60\% centrality interval. Centrality is expressed as percentiles of the inelastic hadronic cross section, with low percentage values corresponding to head-on collisions. The studies are performed separately for particles and antiparticles, and the results are compatible within the statistical uncertainties. Therefore, $v_2$ is the average between results for particles and antiparticles which for the rest of the article will be denoted as \pipm{}, \kapm{}, p+\pbar{}, etc.

This article is organized as follows: the experimental setup is presented in Section~\ref{Sec:ExpSetup}, while the analysis procedure, particle identification (PID), reconstruction methods, and flow measurement techniques are described in Section~\ref{Sec:AnalysisDetails}. Section~\ref{Sec:Systematics} outlines the evaluation of systematic uncertainties. The $v_2$ of  \pipm{}, \kapm{}, p+\pbar{}, \kzero{}, \lmb{}+\almb{}, $\phi$, \X{}+\Ix{}, and \Om{}+\Mo{} and the corresponding flow fluctuations are reported and compared to hydrodynamic calculations in Section~\ref{Sec:Results}. The article concludes with a summary in Section~\ref{Sec:Summary}.

\section{Experimental setup} 
\label{Sec:ExpSetup}

The ALICE detector~\cite{Aamodt:2008zz,Abelev:2014ffa} has been designed to allow detailed physics studies under the extreme conditions created in heavy-ion collisions. ALICE consists of a central barrel that contains several detectors with full or limited azimuthal coverage and a set of forward detectors. The central region is located in a solenoid magnet which generates up to a $0.5$~T field parallel to the beam direction.

The main tracking detectors, positioned in the central barrel, are the Inner Tracking System (\ITS{})~\cite{Aamodt:2008zz} and the Time Projection Chamber (\TPC{})~\cite{Alme:2010ke}. The \ITS{} consists of six layers of silicon detectors employing three different technologies. The two innermost layers are Silicon Pixel Detectors (\SPD{}), followed by two layers of Silicon Drift Detectors (\SDD{}). Finally, the two outermost layers are double-sided Silicon Strip Detectors (\SSD{}). The \SPD{} is also used for event selection and vertex reconstruction. The \TPC{} surrounds the \ITS{} and is also employed for precise tracking of charged particles and for particle identification via the specific energy loss, \dEdx{}. The \dEdx{} is extracted using a truncated-mean procedure, resulting in a \dEdx{} resolution for the 5$\%$ most central Pb--Pb collisions of around 6.5\%, which improves for more peripheral collisions~\cite{Abelev:2014ffa}. The detector provides a separation by at least 2 standard deviations ($\sigma$) for \pipm{}, \kapm{}, and p+\pbar{} at $\pt < 0.7$~\GeVc{} and the possibility to identify particles on a statistical basis for $\pt > 2$~\GeVc{}~\cite{Abelev:2014ffa}. The Time of Flight detector (\TOF{})~\cite{Akindinov:2013tea} is located around the \TPC{} and is used for particle identification by measuring the flight time of particles from the collision point with a resolution of about 80~ps~\cite{Abelev:2014ffa}. The start time for the \TOF{} measurement is provided by the T0 detector with a resolution of about 25 ps~\cite{Abelev:2014ffa,Bondila:2005xy}, two arrays of Cherenkov counters covering the pseudorapidity ranges $-3.3 < \eta < -3.0$ (\TZEROC{}) and $4.6 < \eta < 4.9$ (\TZEROA{}), or from a combinatorial algorithm that uses the particle arrival times at the \TOF{} detector itself~\cite{Abelev:2014ffa, Akindinov:2013tea}. Both methods of estimating the start time are fully efficient for the 60$\%$ most central Pb--Pb collisions. The \TOF{} provides a $3\sigma$ separation between \pipm{}--\kapm{} and \kapm{}--p+\pbar{} up to $\pt = 2.5$~\GeVc{} and $\pt = 4$~\GeVc{}, respectively~\cite{Abelev:2014ffa}. The \ITS{}, \TPC{}, and \TOF{} detectors cover the full azimuth within $|\eta|<0.9$.

In the forward region, two scintillator arrays (\VZERO{})~\cite{Abbas:2013taa} are used for triggering, event selection, and the determination of the collision centrality~\cite{ALICE:2013hur}. The \VZERO{} consists of two systems, the \VZEROC{} and \VZEROA{}, positioned at $-3.7 < \eta < -1.7$ and $2.8 < \eta < 5.1$, respectively. In addition, two tungsten-quartz neutron Zero Degree Calorimeters (ZDCs), installed 112.5 meters from the interaction point on each side, are used for event selection.

More details on the ALICE setup and the performance of the detectors can be found in Refs.~\cite{Aamodt:2008zz,Abelev:2014ffa}.

\section{Analysis procedure}
\label{Sec:AnalysisDetails}

\subsection{Event and track selection} 
\label{Sec:datasample}

The data sample used in this analysis consists of Pb--Pb collisions at $\sqrt{s_{\rm NN}} = 5.02$ TeV recorded by the ALICE detector in the years 2015 and 2018 LHC data-taking campaigns. A minimum bias trigger was provided by requiring signals in both V0A and V0C scintillator arrays. In addition, the sample of semi-central collisions was enhanced by an online selection based on the V0 signal amplitudes. Beam-induced background events (i.e. beam--gas interactions) were removed offline utilizing the V0 and ZDC timing information. Pileup of collisions from different bunch crossings in the TPC was rejected by comparing multiplicity estimates from the V0 detector to those of tracking detectors at midrapidity, exploiting the difference in readout times between the systems. The primary vertex position, determined from tracks reconstructed in the ITS and TPC, was required to be within $\pm$10~cm from the nominal interaction point along the beam direction. These selection criteria were met by approximately 245 million events in the 10--60\% centrality interval. The collision centrality was estimated from the amplitudes of the signals measured in the V0 detector~\cite{ALICE:2013hur}.

Charged-particle tracks, used to measure the $v_2$ of \pipm{}, \kapm{}, p+\pbar{}, $\phi$-mesons, and inclusive charged particles, were reconstructed using the ITS and TPC within $|\eta|<0.8$ and $0.2 < \pt < 10$~\GeVc{}. Each track was required to have a minimum number of 70 TPC space points (out of a maximum of 159) with a $\chi^2$ per TPC space point lower than 4 and at least 2 hits in the ITS with a $\chi^2$ per ITS hit smaller than 36. Moreover, tracks with a distance of closest approach (DCA) larger than 2 cm in the longitudinal direction were rejected. In the transverse plane, a \pt-dependent DCA selection of the form $0.0105 + 0.0350~\pt^{-1.1}$~cm was applied. These selection criteria lead to an efficiency of about 80\% for primary tracks at $\pt > 0.5$~\GeVc{} and contamination from fake tracks (random associations of space points) and secondary charged particles (i.e. particles originating from weak decays, conversions, and secondary hadronic interactions in the detector material) of about 5\% at $\pt \approx 1$~\GeVc{}.

\subsection{Selection of \pipm{}, \kapm{}, and p+$\rm \overline{\mathbf{p}}$} 
\label{Sec:directPID}

Particle identification of \pipm{}, \kapm{}, and p+\pbar{} is performed using the \dEdx{} from the \TPC{} and the time of flight from the \TOF{}~system, if available. The identification is based on the normalised difference between the measured and the expected signal for a given species ($\sigma_{\rm \TPC}$ and $\sigma_{\rm \TOF}$, respectively). It uses the correlation between n$\sigma_{\rm \TPC}$ and n$\sigma_{\rm \TOF}$ in a Bayesian approach~\cite{ALICE:BayesPID}, where the signals converted into probabilities are folded with the expected abundances (priors) of each particle species. The minimal probability threshold has been set to 0.95 for \pipm{} and 0.85 for \kapm{} and p+\pbar{}. In addition, particles are selected by requiring $|$n$\sigma_{\rm \TPC}| < 3$ and $|$n$\sigma_{\rm \TOF}| < 3$ for each species in the whole \pT{} range. This procedure ensures a high purity of the studied sample, thus reducing the uncertainties due to particle misidentification. The resulting purity, estimated using Monte Carlo (MC) simulations, is higher than 95\% for \pipm{} for $0.2<\pt<10$~\GeVc{}, above 80\% for \kapm{} for $0.3<\pt<6$~\GeVc{}, and reaches values larger than 90\% for p+\pbar{} for $0.5<\pt<6$~\GeVc{}.

\subsection{Reconstruction of $\phi$ mesons} 
\label{Sec:recPhi}

The $\phi$ meson is reconstructed in the decay channel $\phi \rightarrow \kap + \kam$ with a branching ratio of 49.2\%~\cite{Zyla:2020zbs}. Its decay products are selected using the same criteria for primary \kapm{} (see Sec.~\ref{Sec:directPID}). The $\phi$ meson yield is obtained from the invariant mass ($M_{\rm \kap \kam}$) reconstructed from all possible \kapm{} pairs after subtracting the combinatorial background evaluated using the like-sign kaon pairs in each \pt{} and centrality interval. The resulting $M_{\rm \kap \kam}$ distribution is parametrised as a sum of a Breit--Wigner (BW) distribution and a third-order polynomial function that accounts for residual contamination within the invariant mass range of 0.99 $< M_{\rm \kap \kam} < $1.07 $\mathrm{GeV}/c^2$. The \pt-differential yield of $\phi$ mesons is extracted by integration of the BW distribution and used for the $v_2$ extraction together with the background yield (see Eq.~\ref{Eq:flowmass}). The procedure of the reconstruction of $\phi$ meson is identical to the previous measurements~\cite{ALICE:2018yph}, while the extraction of $v_{2}\{4\}(p_{\rm T})$ is slightly different and explained in Section~\ref{Sec:FlowMethods}.

\subsection{Reconstruction of \kzero{} and \lmb{}+\almb{}} 
\label{Sec:recV0}

The reconstruction of \kzero{} and \lmb{}+\almb{} is based on identifying their secondary vertices called \Vdecay{}s in the decay channels $\kzero \rightarrow \pip + \pim$ and $\lmb{} \rightarrow {\rm p} + \pim$ ($\almb \rightarrow \pbar + \pip$) with branching ratios of 69.2\% and 63.9\%~\cite{Zyla:2020zbs}, respectively. Selection criteria related to the distinctive V-shaped decay topology and requirements on the characteristics of the daughter particles are applied to suppress the large combinatorial background. The invariant mass is calculated assuming that the daughter particles, identified using the TPC ($|{\rm n\sigma_{TPC}}|<3$) over the entire \pt{} range, are either a \pip{}\pim{} pair or a p\pim{} (\pbar{}\pip{}) pair. The \Vdecay{} candidates are selected with an invariant mass between 0.4 and 0.6~\GeVmass{} for \kzero{} and 1.08 and 1.16~\GeVmass{} for \lmb{}+\almb{}.  The daughter tracks are reconstructed within $|\eta|<0.8$ using the same TPC track quality requirements described in Section~\ref{Sec:datasample} for charged tracks. In addition, the ratio between the number of space points and the number of crossed rows in the TPC is required to be larger than 0.8, the minimum DCA of daughter tracks to the primary vertex is 0.1 cm, and the maximum DCA of daughter tracks to the secondary vertex is 0.5 cm. Only \Vdecay{} candidates produced at a radial distance between 5 and 100 cm from the beam line and with a cosine of the pointing angle (the angle between the line connecting the primary and \Vdecay{} vertices and the \Vdecay{} momentum vector) larger than 0.998 are accepted. To reduce the contamination from \lmb{}+\almb{} and electron--positron pairs coming from $\gamma$ conversions, an additional selection is applied in the Armenteros–Podolanski variables~\cite{armpod} of the \kzero{} candidates, similar to what is done in Ref.~\cite{ALICE:2018yph}. To obtain the \pt-differential yield of \kzero{} and \lmb{}+\almb{}, the invariant mass distributions in various \pt{} intervals are parametrised as a sum of two Gaussian distributions with the same mean and a third-order polynomial function which accounts for residual background. The \kzero{} and \lmb{}+\almb{} yields are extracted by integration of the Gaussian distributions and are not corrected for feed-down from higher mass baryons (e.g. \Xis{}, \Oms{}), but these have a negligible effect on $v_2$~\cite{Abelev:2014pua}.

\subsection{Reconstruction of \X{}+\Ix{} and \Om{}+\Mo{}} 
\label{Sec:recCasc}

The \X{}+\Ix{} and \Om{}+\Mo{} are reconstructed through the cascade topology of the following weak decays: $\X \rightarrow \lmb + \pim$ ($\Ix \rightarrow \almb + \pip$) and $\Om \rightarrow \lmb + \kam$ ($\Mo \rightarrow \almb + \kap$) with branching ratios of 99.9\% and 67.8\%~\cite{Zyla:2020zbs}, respectively, with a subsequent \lmb{} (\almb{}) decay. Candidates are found by applying topological and kinematic criteria first to select the \Vdecay{} with an invariant mass between 1.08 and 1.16~\GeVmass{} and then to match it with one of the remaining secondary tracks. They are selected by requiring the DCA between the \Vdecay{} and the track to be less than 0.3 cm, the cosine of the pointing angle to be at least 0.999 and 0.998 for the cascade and \Vdecay{}, respectively, the DCA between the \Vdecay{} and primary vertex to be larger than \mbox{0.05 cm}, the minimum DCA of \Vdecay{} daughter tracks to the primary vertex to be 0.1 cm, the maximum DCA of \Vdecay{} daughter tracks to be 1.0 cm, and the minimum DCA of the daughter track to the primary vertex to be 0.03 cm. Only \X{}+\Ix{} and \Om{}+\Mo{} candidates produced at a radial distance between 0.9 and 100~cm from the beam line with the same radial distance reported in Section~\ref{Sec:recV0} for \Vdecay{} are accepted. Each of the three daughter tracks is also required to have $\pt>0.15$~\GeVc{} within $|\eta|<0.8$ and to pass the TPC track quality criteria detailed above for charged tracks. In addition, the daughter tracks are checked for compatibility with the pion, kaon, or proton hypotheses by selecting particles with $|{\rm n\sigma_{TPC}}|<3$ for each species. The \pt-differential yield of \X{}+\Ix{} and \Om{}+\Mo{} is obtained by fitting the invariant mass distributions with a sum of two Gaussian distributions with the same mean and a third-order polynomial function that describe the signal and the background, respectively.

\subsection{Flow observables} 
\label{Sec:FlowObservable}
A common approach to study the event-by-event flow fluctuations for a given flow coefficient is by using the two- and multiparticle cumulants~\cite{Borghini:2000sa,Borghini:2001vi}, which have different sensitivities to effects stemming from non-flow and flow fluctuations,
\begin{eqnarray}
v_{\rm n}\{2\} &=& \langle v_{\rm n}^{2} \rangle^{1/2} + \delta_{\rm n}, \\
v_{\rm n}\{4\} &=& \left[2\langle v_{\rm n}^{2} \rangle^{2} - \langle v_{\rm n}^{4} \rangle\right]^{1/4},
\label{Eq:v22v24}
\end{eqnarray}
where $\delta_{\rm n}$ denotes the two-particle non-flow effects.

Assuming that flow fluctuation $\sigma_{\rm v_{n}}$ is relatively small compared to $v_{\rm n}$, which was found to be true for non-central heavy-ion collisions at the LHC~\cite{ATLAS:2013xzf,ALICE:2018rtz,CMS:2017glf}, and also assuming that non-flow effects can be experimentally removed (or largely suppressed, e.g. by using appropriate $\eta$ gaps) in two-particle correlation measurements, the $v_{\rm n}$ can be written as~\cite{Voloshin:2007pc} 
\begin{eqnarray}
v_{\rm n}^{2}\{2\} &=& \langle v_{\rm n} \rangle^{2} + \sigma_{\rm v_{n}}^{2}, \\
v_{\rm n}^{2}\{4\} &\approx& \langle v_{\rm n} \rangle^{2} - \sigma_{\rm v_{n}}^{2},
\label{Eq:v22v24app}
\end{eqnarray}
where $\langle v_{\rm n} \rangle$ and $\sigma_{\rm v_{n}}$ are the mean and fluctuations of the anisotropic flow coefficient, respectively. These two quantities correspond to the first and second moments of the event-by-event $v_{\rm n}$ distribution. The observable $\langle v_{\rm n} \rangle$ is expected to be free from flow fluctuations and to only reflect the true elliptic flow from the flow symmetry plane.

Both $\langle v_{\rm n} \rangle$ and $\sigma_{\rm v_{n}}$ can be calculated using the measured $v_{\rm n}\{2\}$ and $v_{\rm n}\{4\}$ as
\begin{equation}
\langle v_{\rm n} \rangle \approx \sqrt{\frac{ v_{\rm n}^2\{2\} +  v_{\rm n}^2\{4\}} {2}},
\label{eq:meanv}
\end{equation}
\begin{equation}
\sigma_{\rm v_{n}} \approx  \sqrt{\frac{ v_{\rm n}^2\{2\} -  v_{\rm n}^2\{4\}} {2}}.
\label{eq:sigmav}
\end{equation}

Furthermore, the relative flow fluctuations $F(v_{\rm n})$ are defined as
\begin{equation}
F(v_{\rm n}) = \frac{\sigma_{\rm v_{n}}}{\langle v_{\rm n} \rangle}.
\label{Eq:Fvn}
\end{equation}
%

\subsection{Flow extraction methods} 
\label{Sec:FlowMethods}

The measurement of the \pt-differential $v_{\rm n}$ coefficients of identified hadrons is performed using two- and four-particle cumulant method~\cite{Borghini:2000sa}, according to
\begin{equation}\label{vn2}
\mvnTwo{\rm n}(\mpT) = \frac{\mdnTwo{\rm n}}{\sqrt{\mcnTwo{\rm n}}},
\end{equation}
\begin{equation}\label{vn4}
\mvnFour{\rm n}(\mpT) = \frac{-\mdnFour{\rm n}}{ (-\mcnFour{\rm n})^{3/4} }.
\end{equation}
Here $c_{\rm n}\{\rm m\}$ and $d_{\rm n}\{\rm m\}$ are the reference and differential $\textrm{m}$-particle cumulants, respectively, which can be obtained from $\rm m$-particle correlations. For the specific case of two and four particles, they are given by

\begin{equation}
\mcnTwo{\rm n} = \langle\langle 2 \rangle\rangle_{\rm n},
\label{Eq:cn2}
\end{equation}
\begin{equation}
\mdnTwo{\rm n} = \langle\langle 2' \rangle\rangle_{\rm n}, 
\label{Eq:dn2}
\end{equation}
\begin{equation}
\label{Eq:cn4}
\mcnFour{\rm n} = \langle\langle 4 \rangle\rangle_{\rm n} - 2 \, {\langle\langle 2 \rangle\rangle_{\rm n}}^2,
\end{equation}
\begin{equation}
\label{Eq:dn4}
\mdnFour{\rm n} = \langle\langle 4' \rangle\rangle_{\rm n} - 2 \, \langle\langle 2' \rangle\rangle_{\rm n} \cdot \langle\langle 2 \rangle\rangle_{\rm n},
\end{equation}
where $\left<\left< {\rm m} \right>\right>$ and $\left<\left< {\rm m}' \right>\right>$ are the event-averaged reference and differential $\rm m$-particle correlations, respectively. In order to suppress two-particle non-flow correlations, a pseudorapidity gap of $|\Delta \eta| > 0.8$ between the two particles (subevents) is applied when computing the correlations of Eqs.~\ref{Eq:cn2} and~\ref{Eq:dn2}. The relevant flow coefficients will be denoted as $v_2\{2,|\Delta \eta| > 0.8\}$ later in the text.

\begin{figure}[t!]
\vspace{-15pt}
    \begin{center}
    \includegraphics[width = 0.8\textwidth]{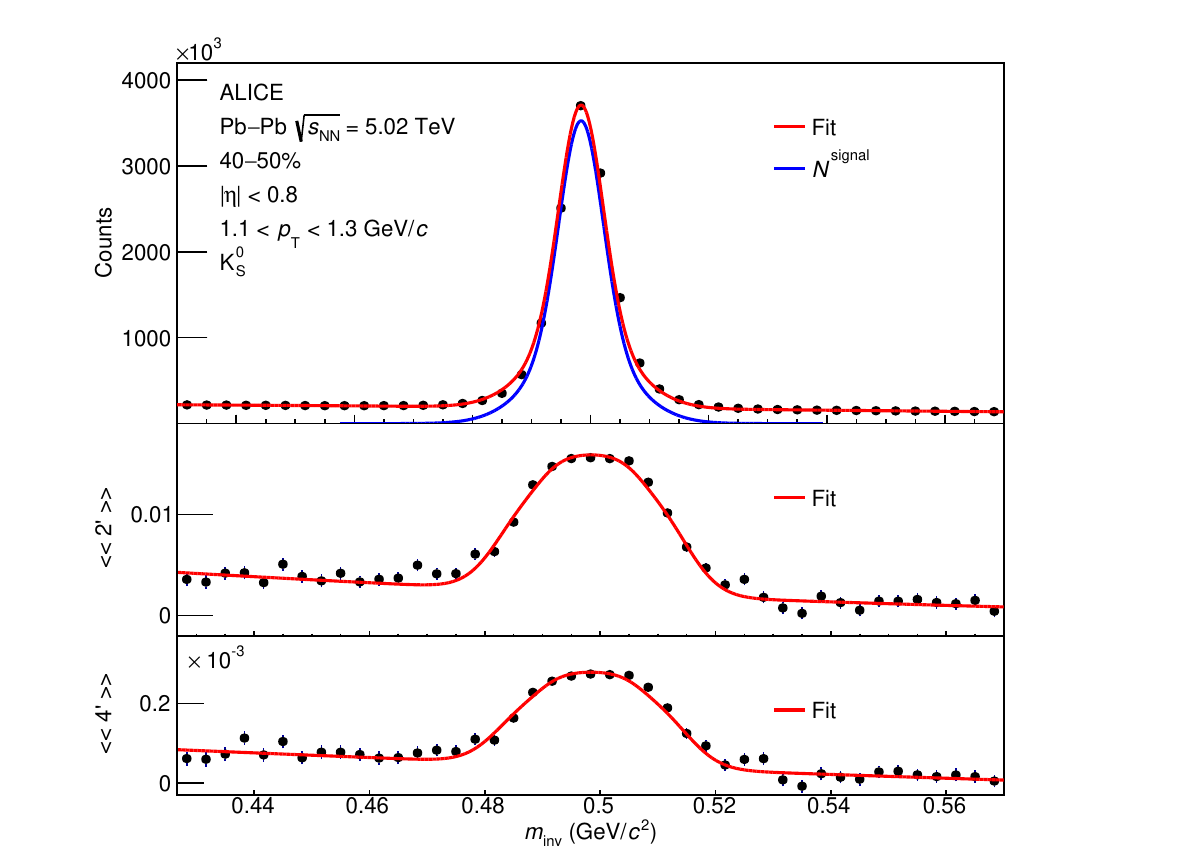}
    \end{center}
    \caption{Simultaneous fits on invariant mass distribution, $\langle \langle 2' \rangle \rangle$, and $\langle \langle 4' \rangle \rangle$ correlations of \Kos~meson at 1.1$<\pt{}<$1.3 GeV/$c$ for centrality 40--50\% in $\sqrt{s_{\mathrm{NN}}} = 5.02$~TeV Pb--Pb collisions.}
    \label{fig:fits}
\end{figure}

The multiparticle correlation technique with the generic framework implementation~\cite{Bilandzic:2013kga} allows for correcting for detector inefficiencies and non-uniformities in the azimuthal particle distribution using weights. Using this method, one can measure the \pt-differential flow with two- and four-particle cumulants of inclusive charged hadrons, \pipm{}, \kapm{}, and p+\pbar{} for each centrality percentile. For \kzero{}, \lmb{}+\almb{}, $\phi$, \X{}+\Ix{}, and \Om{}+\Mo{}, the identification on a particle-by-particle basis is not possible. Besides a signal component (true particles), the candidates contain a non-negligible combinatorial background. 
Considering the fact that the corresponding \pt{}-differential multiparticle cumulants of both signal and background might not be easily decomposed into individual contributions, the invariant mass method~\cite{Borghini:2004ra} was developed exploiting the additivity of correlation. This method has been used in previous anisotropic flow measurements for reconstructed particles with the two-particle correlation method ~\cite{Abelev:2014pua,ALICE:2018yph,ALICE:2019xkq} and is also used in this analysis. The two- and four-particle correlations are measured as a function of both invariant mass and candidate \pt{}. The relation between the signal and background components is given for each \pt{} interval by
\begin{equation} 
\label{Eq:flowmass}
\avcor{{\rm m}'}{\rm n}^\mathrm{total}(m_\mathrm{inv}) = f^\mathrm{signal}(m_\mathrm{inv}) \, \avcor{{\rm m}'}{\rm n}^\mathrm{signal}\ +  f^\mathrm{bg} \, \avcor{{\rm m}'}{\rm n}^\mathrm{bg}(m_\mathrm{inv}),
\end{equation}
where the total $\rm m$-particle correlation $\avcor{{\rm m}'}{\rm n}^\mathrm{total}$ can be regarded as the sum of the $\rm m$-particle correlations of signal particles $\avcor{{\rm m}'}{\rm n}^\mathrm{signal}$ and the correlations of the background $\avcor{{\rm m}'}{\rm n}^\mathrm{bg}$. Here the signal function is weighted by its corresponding fraction $f^\mathrm{signal}$ defined as
\begin{equation} \label{Eq:flowmass-fraction}
f^\mathrm{signal}(m_\mathrm{inv}) = \frac{N^\mathrm{signal}(m_\mathrm{inv}) }{N^\mathrm{signal}(m_\mathrm{inv})  + N^\mathrm{bg}(m_\mathrm{inv}) },
\end{equation}
where $N^\mathrm{signal}(m_\mathrm{inv})$ and $N^\mathrm{bg}(m_\mathrm{inv})$ are signal and background yields, respectively. Correspondingly, the weight of the background function is determined with $f^\mathrm{bg} = 1 - f^\mathrm{sig}$. Both $N^\mathrm{signal}(m_\mathrm{inv})$ and $N^\mathrm{bg}(m_\mathrm{inv})$ are obtained from the invariant mass distribution of \kzero{}, \lmb{}+\almb{}, $\phi$, \X{}+\Ix{}, and \Om{}+\Mo{} for each \pt{} interval and centrality class following the procedures outlined in Sections~\ref{Sec:recPhi},~\ref{Sec:recV0}, and~\ref{Sec:recCasc}. After measuring $\avcor{{\rm m}'}{\rm n}^\mathrm{total}(m_\mathrm{inv})$ via multiparticle correlations, the $\avcor{{\rm m}'}{\rm n}^\mathrm{signal}$ can be determined for a given centrality class and \pt{} interval using a simultaneous fit to the $\avcor{{\rm m}'}{\rm n}^\mathrm{total}(m_\mathrm{inv})$ and $N^\mathrm{total}(m_\mathrm{inv})$ distributions, where $\avcor{{\rm m}'}{\rm n}^\mathrm{bg}(m_\mathrm{inv})$ is parametrised as a first-order polynomial function. An example of such a procedure with distributions of $m_\mathrm{inv}, \langle \langle 2' \rangle \rangle$, and $\langle \langle 4' \rangle \rangle$ together with fit functions is shown in Fig.~\ref{fig:fits} for \Kos~in  1.1$<\pt{}<$1.3 GeV/$c$ and centrality 40--50\%. The result of the fit makes it possible to calculate the corresponding $v_{2}\{2\}(p_{\rm T})$ and $v_{2}\{4\}(p_{\rm T})$ using Eqs.~\ref{vn2} and~\ref{vn4}. 

\section{Systematic uncertainties} 
\label{Sec:Systematics}

The systematic uncertainties were evaluated by varying the selection criteria for each particle species, in every \pt{} range and centrality interval, such as track quality criteria for \pipm{}, \kapm{}, and p+\pbar{} or topological reconstruction requirements for $\phi$, \kzero{}, \lmb{}+\almb{}, \X{}+\Ix{}, and \Om{}+\Mo{}. Only statistically significant differences between the nominal data points and the systematic variations, where significance is evaluated based on the recommendations in Ref.~\cite{Barlow:2002yb}, were assigned as systematic uncertainties. The uncertainties from the independent sources were added in quadrature to obtain the final systematic uncertainties on the measurements. For each particle species, a \pt-dependent systematic uncertainty is reported for $\pt<3$~\GeVc{}, while a \pt-independent average uncertainty is assigned at higher transverse momenta to suppress statistical fluctuations. Tables~\ref{tab:syst_direct},~\ref{tab:syst_reconstructed}, and~\ref{tab:syst_xiOmega} summarise the minimum and maximum values of the relative systematic uncertainties per particle species for all \pt{} and centrality ranges.

For the event selection criteria, the primary vertex position along the beam line was varied from 10 cm to 8 cm, and the centrality determination was changed from energy deposition in the V0 scintillator to the number of hits in the first layer of the ITS. Additionally, for  \pipm{}, \kapm{}, and p+\pbar{}, the event sample was separated based on the polarity of the ALICE solenoid, and the two sub-samples were studied independently.

Systematic uncertainties arising from the selection criteria imposed at the track level were investigated by requiring only tracks that have at least three hits per track in the ITS complemented by tracks without hits in the first two layers of the ITS (in which case the primary interaction vertex is used as an additional constraint for the momentum determination). In addition, systematics uncertainties in the measurements were investigated by increasing the minimum number of TPC space points from 70 to 90, and by varying the DCA from the strict \pt-dependent selection to 0.15~cm in the transverse plane and from 2~cm to 0.2~cm in the longitudinal direction, in order to estimate the impact of secondary particles. These variations are referred as ``tracking mode" in Tables~\ref{tab:syst_direct} and~\ref{tab:syst_xiOmega}. Furthermore, the minimal probability threshold in the Bayesian particle identification was increased from 0.95 to 0.98 for \pipm{} and from 0.85 to 0.9 for \kapm{} and p+\pbar{}.

\begin{table}
	\caption{The minimum and maximum values of the relative systematic uncertainties from each individual source for \pipm{}, \kapm{}, and p+\pbar{}. Percentage ranges are given to account for variations with \pt{} and centrality. 
	}
\centering
		\begin{tabular}[!ht]{|l||c|c|c||c|c|c|c|}
		\hline
		& \multicolumn{3}{c||}{$v_2\{2,|\Delta\eta|>0.8\}$} & \multicolumn{3}{c|}{$v_2\{4\}$}  \\
		\hline
		Uncertainty source & \pipm{} & \kapm{} & p+\pbar{} & \pipm{} & \kapm{} & p+\pbar{} \\
		\hline
		\hline
		Centrality estimator &  0--1\% & 0--1\% & 0--1\% & 0--1\% & 0--1\% & 0--1\% \\
		\hline
		Magnetic field polarity & 0--1\% & 0--1\% & 0--1\% & 0--1\% & 1--3\% & 0--3\% \\
		\hline
		Tracking mode & 0--2\% & 0--5\% & 0--5\% & 0--1\% & 0--1\% & 0--2\% \\
		\hline
		Bayesian particle identification & 0--5\% & 0--5\% & 0--4\% & 0--5\% & 0--4\% & 0--4\% \\
		\hline
	\end{tabular}
	\label{tab:syst_direct}
\end{table}

For \kzero{} and \lmb{}+\almb{}, the topological requirements on the \Vdecay{}s themselves were varied by changing the maximum DCA of the \Vdecay daughter tracks to the secondary vertex from 0.5 cm to 0.3 cm and the minimum DCA of the \Vdecay daughter tracks to the primary vertex from 0.1 cm to 0.3 cm. In addition, the minimum radial distance to the primary vertex at which the \Vdecay can be produced was changed from 5~cm to 1~cm and 10~cm. The selection criteria imposed on the daughter tracks were varied by increasing the minimum number of TPC space points from 70 to 90, requiring the ratio between the number of space points and the number of crossed rows in the TPC to be larger than 0.9 or 1.0 instead of 0.8 (denoted as ``track quality" in Table~\ref{tab:syst_reconstructed}), and requiring a minimum \pt{} of 0.2~\GeVc{}. Finally, the strategy for reconstructing \Vdecay was changed from online, where \Vdecay{}s were reconstructed during the track fitting procedure, to offline, which took place after all the tracks have been reconstructed.

\begin{table}
\caption{The minimum and maximum values of the relative systematic uncertainties from each individual source for \kzero{}, \lmb{}+\almb{}, and $\phi$ meson. Percentage ranges are given to account for all \pt{} and centrality intervals. The fields marked as "negl." (negligible) denote that the uncertainties were tested but are not statistically significant.
}
	\centering
	\begin{tabular}[!ht]{|l||c|c|c||c|c|c|c|}
		\hline
		& \multicolumn{3}{c||}{$v_2\{2,|\Delta\eta|>0.8\}$} & \multicolumn{3}{c|}{$v_2\{4\}$}\\
		\hline
		 Uncertainty source & \kzero{} & \lmb{}+\almb{} & $\phi$ meson & \kzero{} & \lmb{}+\almb{} & $\phi$ meson \\
		\hline
		\hline
	     Centrality estimator & 0--1\% & 0--1\% & 2--3\% & 1--3\% & 1\% & 2--5\% \\
		\hline
	    Track quality & 1--3\% & 0--2\% & negl. & 1--5\% & 0--4\% & negl. \\
		\hline
		Background fit function & negl. & negl. & negl. & negl. & negl. & negl. \\
		\hline
	    Signal fit function & negl. & negl. & negl. & negl. & negl. & negl. \\
		\hline
		\Vdecay{} finding strategy & 0--1\% & 0--1\%  & -- & 0--2\% & 0--1\% & -- \\
		\hline
		DCA decay products to primary vertex & 0--1\% & 0--1\% & -- & 1\% & 0--1\% & --  \\
		\hline
		DCA between decay products & 0--1\% & 0--1\% & -- & 0--1\% & 0--1\% & -- \\
		\hline
		Minimum \pt{} of daughter tracks & negl. & 0--1\% & -- & 0--1\% & 0--1\% & --  \\
		\hline
		Decay vertex (radial position 1 cm) & 1\% & 0--1\% & -- & 2\% & 0--1\% & --   \\
		\hline
		Decay vertex (radial position 10 cm) & 1--2\% & 1\% & -- & 1--5\% & 1\% &  \\		
	    \hline
	\end{tabular}
	\label{tab:syst_reconstructed}
\end{table}

The \X{}+\Ix{} and \Om{}+\Mo{} finding criteria were varied by changing the maximum DCA between the \Vdecay{} and bachelor track from 0.3 cm to 0.25 cm, increasing the minimum DCA between the \Vdecay{} and primary vertex from 0.05 cm to 0.06 cm, decreasing the minimum DCA of \Vdecay{} bachelor tracks to the primary vertex from 0.1 cm to 0.08 cm. In addition, the criteria were changed by requiring the maximum DCA of \Vdecay{} bachelor tracks to be 0.8 cm instead of 1.0 cm, increasing the minimum DCA of the bachelor track to the primary vertex from 0.03 cm to 0.035 cm, and decreasing the minimum value of the cosine of the pointing angle for the cascade from 0.999 to 0.995. For the \Vdecay, the invariant mass range was changed from (1.08--1.16)~\GeVmass{} to (1.10--1.14 )~\GeVmass{}, while the minimum radial distance was varied from 5~cm to 1~cm and 10~cm. A minimum value of 0.995 instead of 0.998 was used for the cosine of the pointing angle. For each of the three daughter tracks, the PID criterion was varied from $|{\rm n\sigma_{TPC}}|<3$ to $|{\rm n\sigma_{TPC}}|<2$ and the minimum \pt{} was increased to 0.2~\GeVc{}.

An additional contribution from fitting parameter variations was studied for all the reconstructed particles, following the same approach used in previous works~\cite{Abelev:2014pua,ALICE:2019xkq}. The resulting systematic uncertainties, negligible for \kzero{}, \lmb{}+\almb{}, and $\phi$ mesons, but significant for \X{}+\Ix{} and \Om{}+\Mo{}, are summarized in Table~\ref{tab:syst_xiOmega}.

\begin{table}
\caption{The minimum and maximum values of the relative systematic uncertainties from each individual source for \X{}+\Ix{} and \Om{}+\Mo{}. Percentage ranges are given to account for all \pt{} and centrality intervals. The fields marked as "negl." (negligible) denote that the uncertainties were tested but are not statistically significant.
}
	\centering
	\begin{tabular}[!ht]{|l||c|c||c|c|c|}
		\hline
		& \multicolumn{2}{c||}{$v_2\{2,|\Delta\eta|>0.8\}$} & \multicolumn{2}{c|}{$v_2\{4\}$}\\
		\hline
		 Uncertainty source  & \X{}+\Ix{} & \Om{}+\Mo{} &  \X{}+\Ix{} & \Om{}+\Mo{} \\
		\hline
		\hline
		 Centrality estimator  & 0--1\% & 0--4\% & 1\% & 2--6\% \\
		\hline
		 Number of TPC space points & 0--1\% & 0--2\% & negl. & 0--2\%  \\
		\hline
		 Tracking mode for bachelor track & negl. & 0--1\% & negl. & 1--3\%  \\
		 \hline
		 Particle identification of decay products & negl. & 1--2\% & negl. & 1-- 3\% \\
		 \hline
		 \Vdecay{} invariant mass range & negl. & 1--2\% & negl. & 1-- 3\% \\
		\hline 
		DCA between \Vdecay{} decay products &  0--1\% & 1--2\% & 0--1\%& 0-- 1\% \\		
		\hline
		DCA between \Vdecay{} and primary vertex & negl. & 0--1\% & 0--1\%& 1-- 2\% \\
		\hline
		Decay vertex of \Vdecay{} (radial position 10 cm) & negl. & 0--1\% & negl. & 0--1\% \\
		\hline
		Background fit function & negl. & negl. & 0--1\% & negl. \\
		\hline
	    Signal fit function & 1--3\% & 1--2\%  & 1--3\% & 1-- 2\% \\
		\hline
	\end{tabular}
	\label{tab:syst_xiOmega}
\end{table}

\section{Results} 
\label{Sec:Results}

\begin{figure}[t!]
    \begin{center}
    \includegraphics[width = \textwidth]{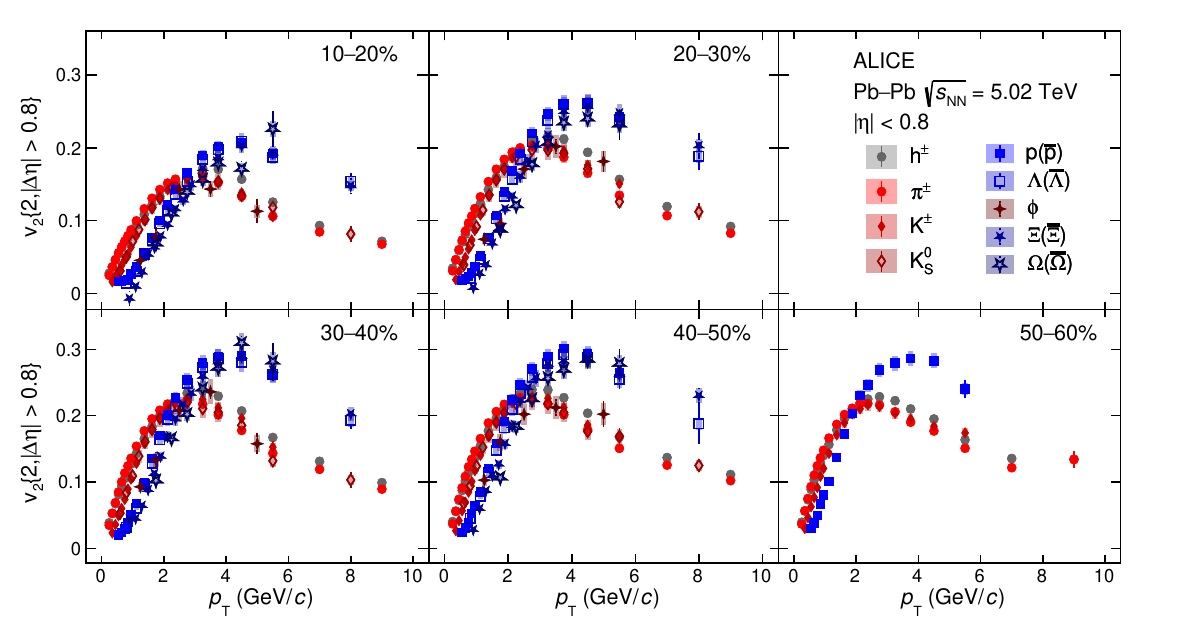}
    \end{center}
    \caption{The \pt-differential $v_2$ measured with two-particle 
    correlations with a pseudorapidity gap of $|\Delta \eta| > 0.8$ 
    for different particle species and centralities in Pb--Pb collisions at $\sqrt{s_{\mathrm{NN}}} = 5.02$~TeV. The vertical error bars and the filled boxes represent statistical and systematic uncertainties, respectively.}
    \label{fig:v22}
\end{figure}
\begin{figure}[t!]
    \begin{center}
    \includegraphics[width = \textwidth]{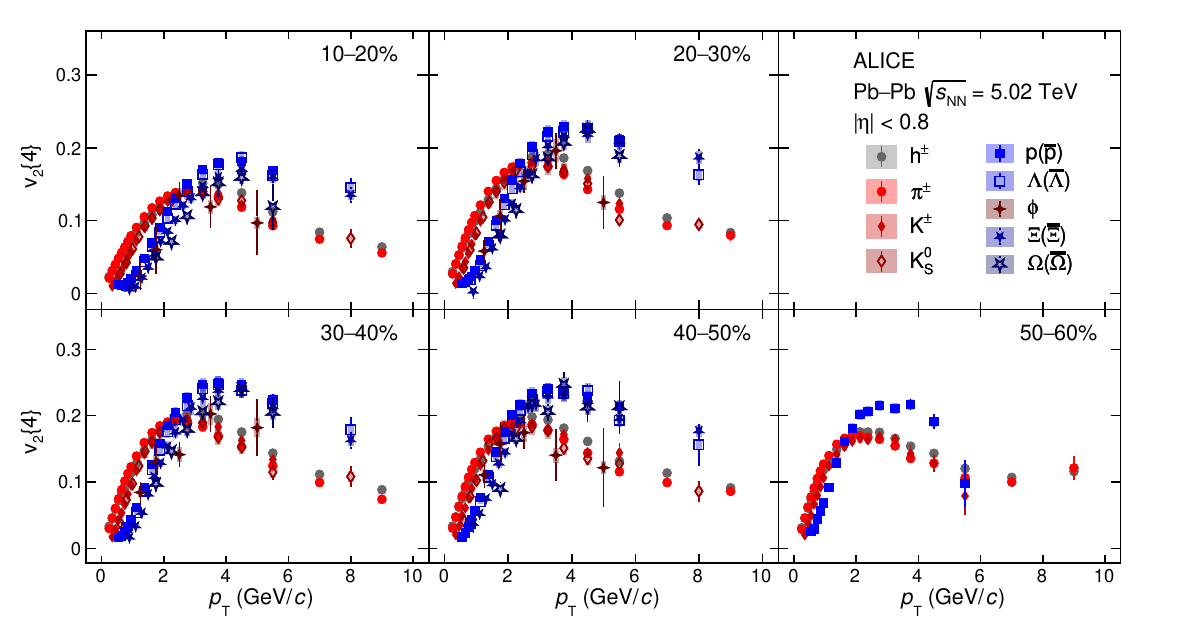}
    \end{center}
    \caption{The \pT-differential $v_2$ measured with four-particle 
    cumulants ($v_2\{4\}$) for different particle species and centralities in Pb--Pb collisions at $\sqrt{s_{\mathrm{NN}}} = 5.02$~TeV. The vertical error bars and the filled boxes represent statistical and systematic uncertainties, respectively.}
    \label{fig:v24}
\end{figure}

In this section, the results for the \pt-differential $v_2$ and its relative fluctuations measured in various collision centrality intervals of Pb--Pb collisions at $\sqrt{s_{\mathrm{NN}}} = 5.02$~TeV are presented. The $v_2$ measured with two- and four-particle cumulants and the corresponding results for flow fluctuations for different particle species are reported in Sec.~\ref{SubSection:MassNCQ} and Sec.~\ref{SubSection:Fluctuations}, respectively. In Section~\ref{SubSection:Models}, the experimental data are compared with a state-of-the-art hydrodynamic model calculations, namely, the coupled linear Boltzmann transport (CoLBT)~\cite{Chen:2020tbl}, that applies $\rm T_{R}ENTo$ initial state model and incorporates the bulk expansion of the medium with a specific shear viscosity $\eta/s$ $=$ 0.10 and interactions of energetic partons with it, as well as a coalescence mechanism for particle production. Note that the same data will be shown in different representations to highlight the various physics implications of the measurements. The data points will be drawn together with their statistical and systematic uncertainties represented by the error bars and shaded boxes around each marker, respectively. 

\subsection{Mass ordering and scaling properties}
\label{SubSection:MassNCQ}

Figure~\ref{fig:v22} presents the \pt-differential $v_2\{2,|\Delta \eta| > 0.8\}$ measurements for unidentified charged hadrons ($\rm h^{\pm}$) as well as for \pipm{}, \kapm{}, p+\pbar{}, \kzero{}, \lmb{}+\almb{}, $\phi$, \X{}+\Ix{}, and \Om{}+\Mo{} from Pb--Pb collisions at $\sqrt{s_{\mathrm{NN}}} = 5.02$~TeV. 
The five panels show different centrality intervals, with the most central and the most peripheral, i.e. 10--20\% and 50--60\%, drawn in the top left and bottom right, respectively. This analysis profits from the data samples collected by ALICE in 2015 and 2018 which allow extending the previous results from two-particle correlations~\cite{Abelev:2014pua,ALICE:2012vgf,ALICE:2016cti,ALICE:2019xkq} to higher-order cumulants. Figure~\ref{fig:v24} presents the first \pt-differential $v_2$ measurements using four-particle cumulants (i.e. $v_2\{4\}$) for the same particle species as reported in Fig.~\ref{fig:v22} from Pb--Pb collisions at $\sqrt{s_{\mathrm{NN}}} = 5.02$~TeV. In both cases, similar features of the \pt-differential measurements as reported and discussed in detail in Refs.~\cite{Abelev:2014pua,ALICE:2016cti,ALICE:2018yph,ALICE:2019xkq} are confirmed. The progressive increase of $v_2$ with the centrality of the collision for a given \pt~interval illustrates the final-state anisotropy that originates from the initial-state ellipsoidal geometry in non-central collisions, quantified by the spatial eccentricity $\epsilon_2$. Furthermore, both the effects known in the literature as mass ordering and the meson--baryon particle type grouping are present in these new measurements. The former originates from the radial flow of the system, while the latter is explained in a dynamical picture where flow develops at the partonic level followed by quark coalescence into hadrons~\cite{Molnar:2003ff}.

\begin{figure}[t!]
    \begin{center}
    \includegraphics[width = \textwidth]{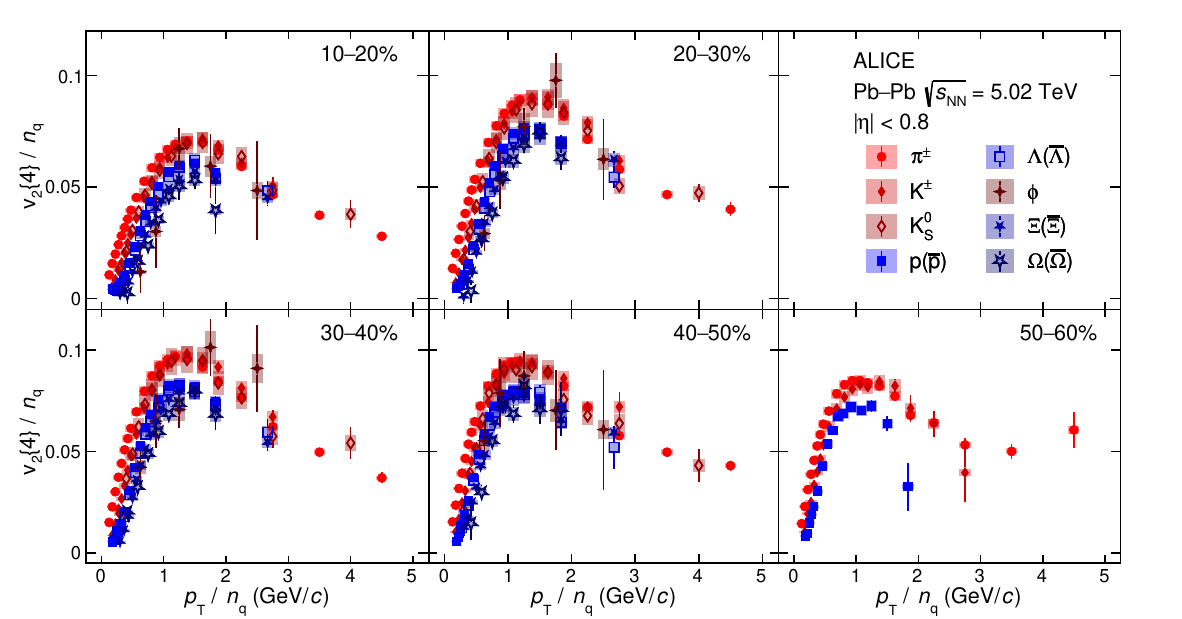}
    \end{center}
    \caption{The dependence of $v_2\{4\}/n_{\rm q}$ on $\pt/n_{\rm q}$, where $n_{\rm q}$ is the number of constituents quarks, for different particle species and centralities in Pb--Pb collisions at $\sqrt{s_{\mathrm{NN}}} = 5.02$~TeV. The vertical error bars and the filled boxes represent statistical and systematic uncertainties, respectively.}
    \label{fig:v24NCQ}
\end{figure}

\begin{figure}[t!]
    \begin{center}
    \includegraphics[width = \textwidth]{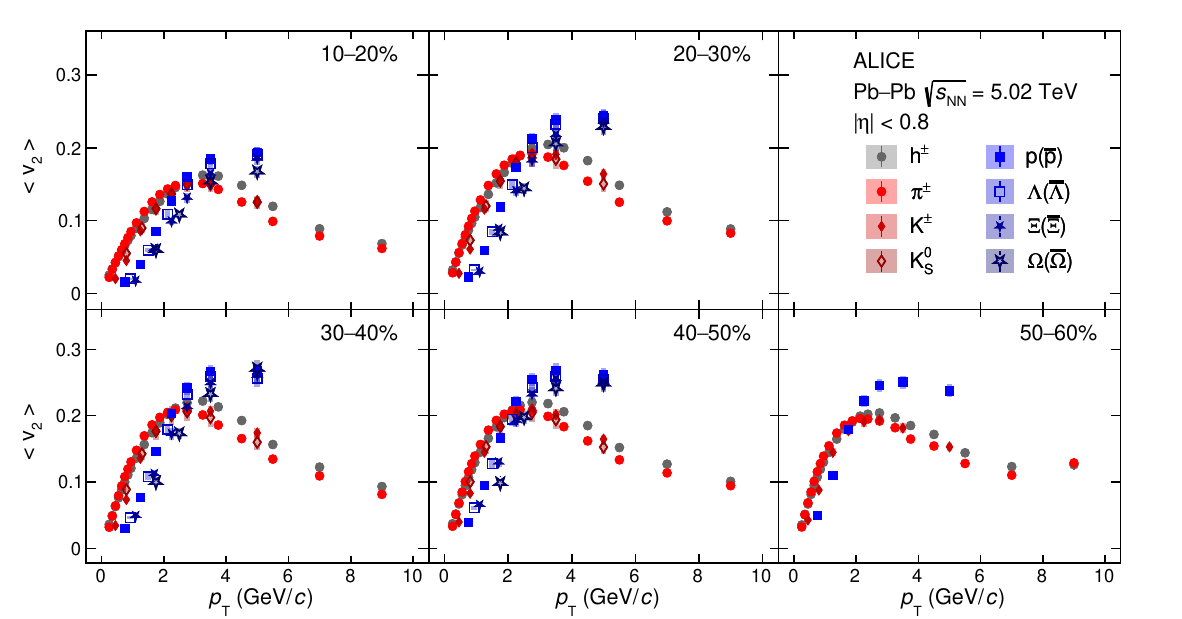}
    \end{center}
    \caption{The dependence of the mean value of $v_2$ 
    ($\langle v_2 \rangle$) on \pt{} for different 
    particle species and centralities in Pb--Pb collisions at 
    $\sqrt{s_{\mathrm{NN}}} = 5.02$~TeV. The vertical error bars and the filled boxes represent statistical and systematic uncertainties, respectively.}
    \label{fig:MeanV2}
\end{figure}

The meson--baryon grouping is generally attributed to hadron production via coalescence in the intermediate \pt{} region~\cite{Molnar:2003ff}, where the direct contribution from hydrodynamic expansion may no longer be dominant and the path-length dependence of energy loss might not play a significant role yet~\cite{Zhao:2021vmu}. This grouping is further investigated using the number of constituent quarks (NCQ) scaling, similarly to what was done in Refs.~\cite{Abelev:2014pua,ALICE:2016cti,ALICE:2018yph,ALICE:2019xkq}. The values of $v_2\{4\}/n_{\rm q}$ reported in Fig.~\ref{fig:v24NCQ} confirm that the scaling, if it holds at all, is only approximate. The contributions of  $v_2\{4\}$ from different sources will be further discussed in section~\ref{SubSection:Models}, where detailed comparisons with theoretical model calculations will be shown.

\subsection{Results on flow fluctuations}
\label{SubSection:Fluctuations}

\begin{figure}[t!]
    \begin{center}
    \includegraphics[width = \textwidth]{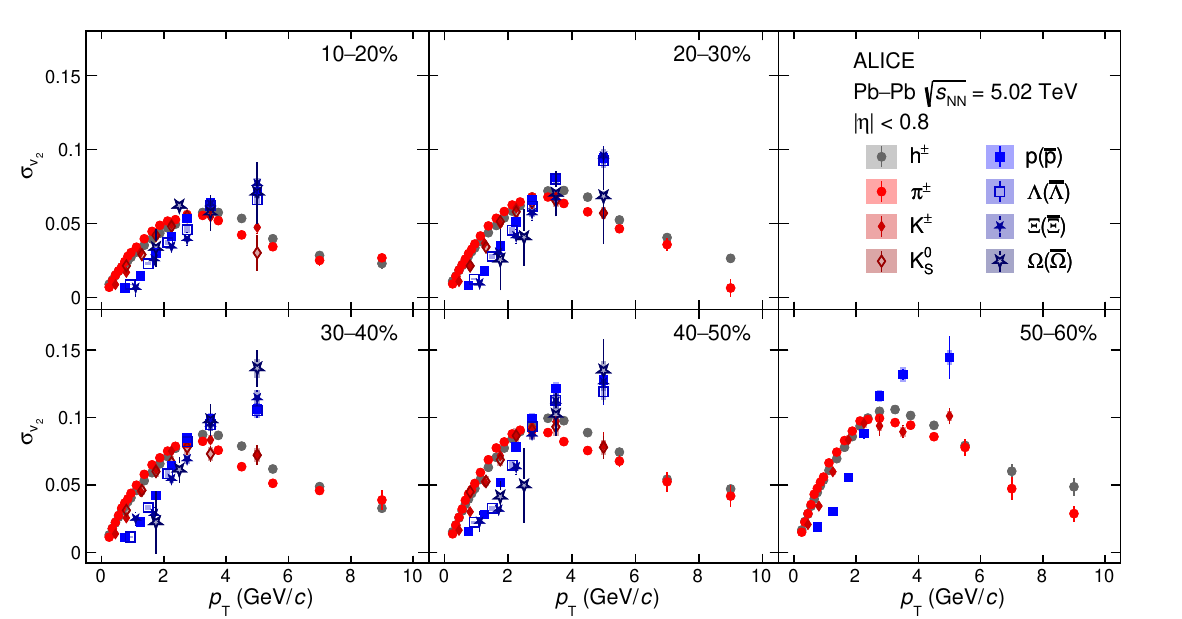}
    \end{center}
    \caption{The \pt{} dependence of the standard deviation of $v_2$ ($\sigma_{\rm v_2}$) for different particle species and centralities in Pb--Pb collisions at $\sqrt{s_{\mathrm{NN}}} = 5.02$~TeV. The vertical error bars and the filled boxes represent statistical and systematic uncertainties, respectively.}
    \label{fig:SigmaV2}
\end{figure}
\begin{figure}[t!]
    \begin{center}
    \includegraphics[width = \textwidth]{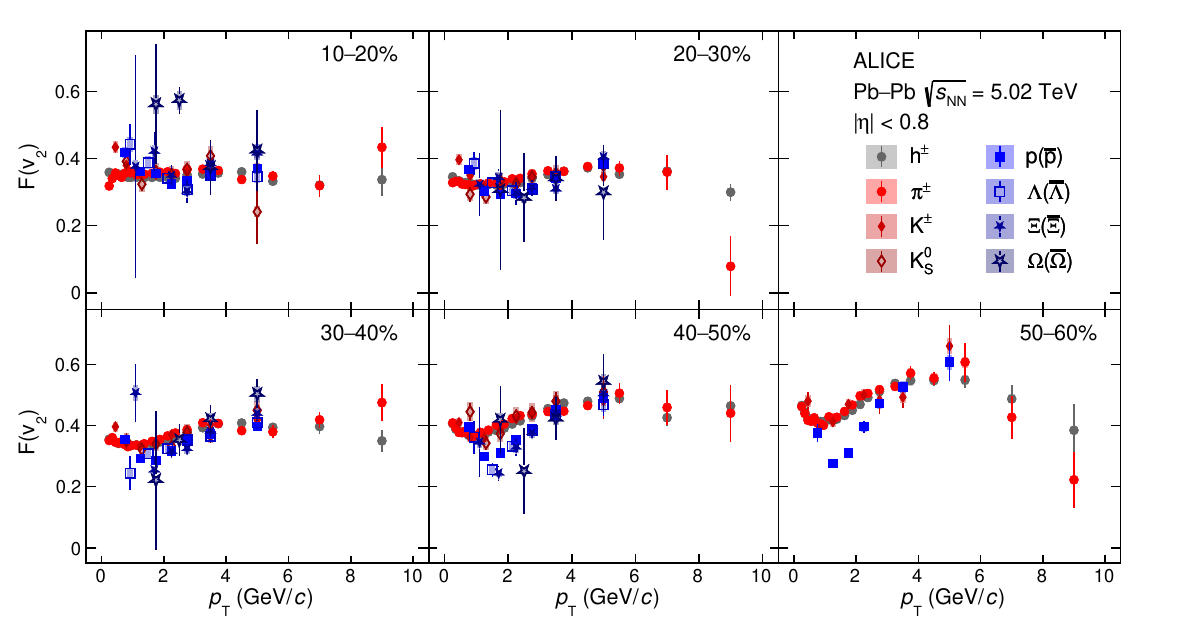}
    \end{center}
    \caption{The relative elliptic flow fluctuations ($F(v_2)$) as a function of \pt{} for different particle species and centralities in Pb--Pb collisions at $\sqrt{s_{\mathrm{NN}}} = 5.02$~TeV. The vertical error bars and the filled boxes represent statistical and systematic uncertainties, respectively.}
    \label{fig:FlucV2}
\end{figure}

The measurements of $v_2$ with two- and four-particle cumulants provide the first opportunity to investigate the first moments of the $v_2$ distribution for different particle species. Figure~\ref{fig:MeanV2} presents the mean value of $v_2$, denoted as $\langle v_2 \rangle$, as a function of \pt{} for the same combination of hadrons and centrality intervals as in the previous figures. Assuming that the non-flow contribution in $v_2^2\{2,|\Delta \eta| > 0.8\}$ is negligible~\cite{ALICE:2017fcd,ALICE:2017lyf}, this mean value is calculated according to Eq.~\ref{eq:meanv} by replacing $v_2^2\{2\}$ with $v_2^2\{2,|\Delta \eta| > 0.8\}$ measurement. 

Figure~\ref{fig:SigmaV2} presents the transverse momentum dependence of the second moment of the $v_2$ distribution, i.e. the standard deviation $\sigma_{\rm v_2}$, measured for the first time for different particle species. As in the previous case, assuming negligible non-flow contribution in $v_2^2\{2,|\Delta \eta| > 0.8\}$, $\sigma_{\rm v_2}$ is approximated according to Eq.~\ref{eq:sigmav}. The data points of both $\langle v_2 \rangle$ and $\sigma_{\rm v_2}$ show, as expected, the same qualitative features as in the previous cases of Figs.~\ref{fig:v22} and~\ref{fig:v24}: namely the mass ordering developing at low values of \pt{} and the particle type grouping that is evident at higher \pt{}.

Combining $\langle v_2 \rangle$ and $\sigma_{\rm v_2}$, one can quantify the relative $v_2$ fluctuations ($F(v_2)$) according to Eq.~\ref{Eq:Fvn}. This quantity is displayed in Fig.~\ref{fig:FlucV2} as a 
function of \pt{} and centrality intervals for the 
various particle species presented in this article. It can be seen 
that for central events, there is no 
significant \pt{} or particle species dependence. However, for more 
peripheral collisions, and in particular starting from the interval 
30--40\% and above, the data points exhibit a non-monotonic transverse momentum dependence, with a minimum in $F(v_2)$ that lies at higher values of \pt{} for baryons than for mesons. In addition, the $F(v_2)$ for baryons in $1 < \pt < 3$~\GeVc{} is systematically lower than for mesons. Interestingly, the momentum region where this apparent particle type grouping develops for $F(v_2)$ does not coincide with the region where a similar grouping is reported for measurements of $v_2$. This could point to a different origin for these two observations. For $\pt > 3$~\GeVc{}, all data points converge into a universal band within the uncertainties. The origin of this characteristic behaviour of $F(v_2)$ is studied using hydrodynamical models in the following section.

\begin{figure}[tb]
    \begin{center}
    \includegraphics[width = \textwidth]{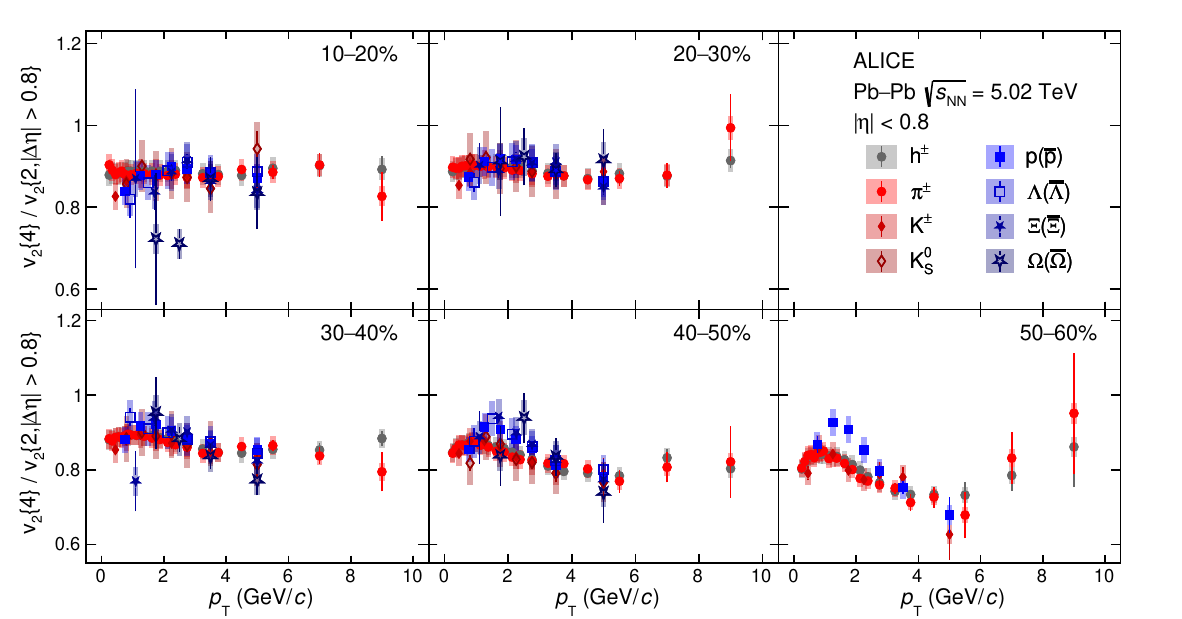}
    \end{center}
    \caption{The ratio $v_2\{4\}/v_2\{2,|\Delta \eta| > 0.8\}$ as a 
    function of \pt{} for different particle species and centralities in Pb--Pb collisions at $\sqrt{s_{\mathrm{NN}}} = 5.02$~TeV. The vertical error bars and the filled boxes represent statistical and systematic uncertainties, respectively.}
    \label{fig:Ratio}
\end{figure}

Finally, to further study the nature of flow fluctuations, Fig.~\ref{fig:Ratio} presents the \pt-dependence of the ratio $v_2\{4\}/v_2\{2,|\Delta \eta| > 0.8\}$. This ratio is expected to be sensitive to the fluctuations within the picture of initial state models. Within these models, the spatial eccentricity $\epsilon_2$ fluctuates from event to event. These fluctuations are transferred through the low viscosity QGP to the final state and are imprinted in how $v_2$ fluctuates. Since $v_2 \propto \epsilon_2$, the ratio $v_2\{4\}/v_2\{2,|\Delta \eta| > 0.8\}$ is expected to reflect the ratio between  $\epsilon_2\{4\}$ and $\epsilon_2\{2\}$, which have positive and negative contributions from the initial eccentricity fluctuations, and thus can provide strong constraints on initial state models. It can be seen that for central collisions, this ratio does not exhibit any significant \pt{} or particle species dependence. Starting from the 20--30\% centrality interval, however, a decrease in the ratio can be seen between 1 and 5~\GeVc{}. It becomes progressively more pronounced for more peripheral events. In addition, starting from the 30--40\% centrality interval, and similar to the picture that develops for $F(v_2)$, the data points indicate a particle type grouping, with the values for baryons being systematically larger than the ones for mesons. This apparent dependence on particle species highlights that final state effects play a significant role in these observables. 
%

\subsection{Comparison with models}
\label{SubSection:Models}

\begin{figure}[tb]
    \begin{center}
    \includegraphics[width = 0.49\textwidth,trim={0 0 1cm 1cm}, clip]{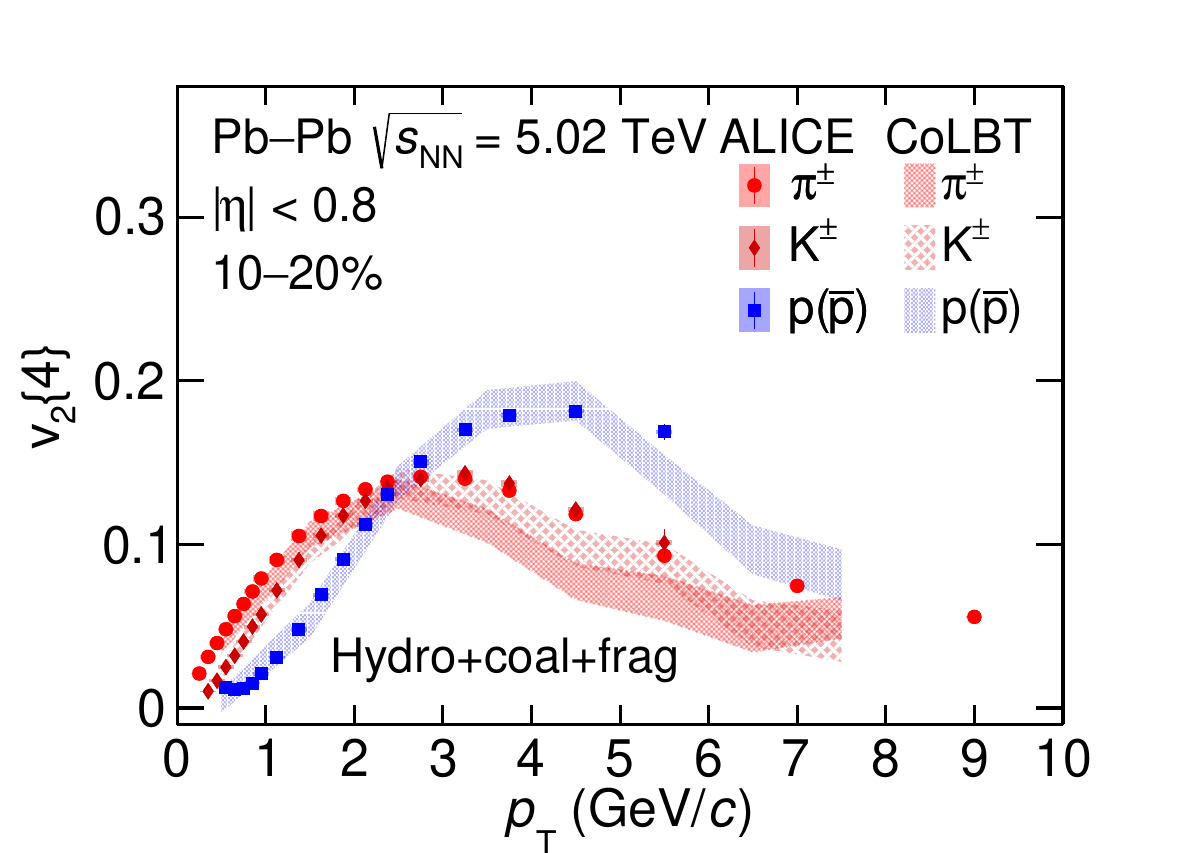}
    \includegraphics[width = 0.49\textwidth,trim={0 0 1cm 1cm}, clip]{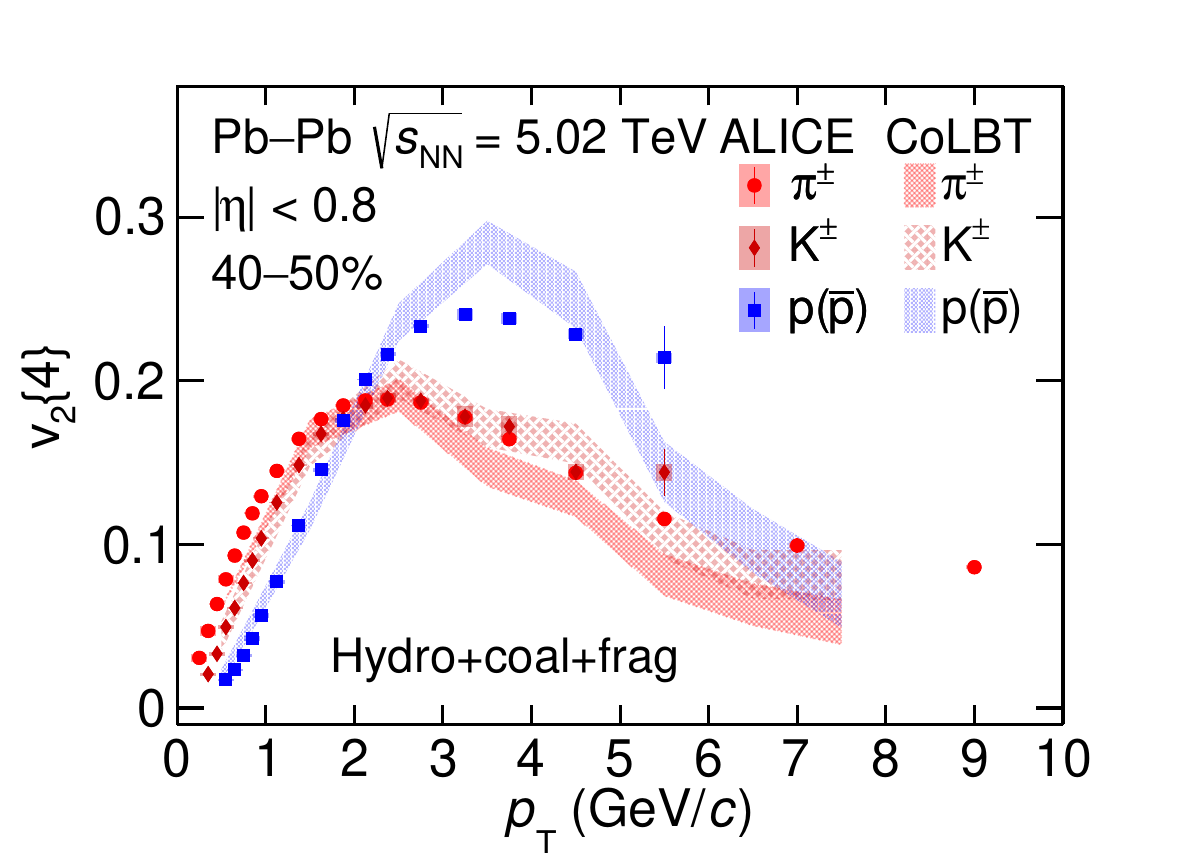}
    \end{center}
    \caption{The \pt-differential $v_2\{4\}$ for \pipm{}, \kapm{}, and p+\pbar{} measured in Pb--Pb collisions at $\sqrt{s_{\mathrm{NN}}} = 5.02$~TeV compared with expectations of the same quantity from the CoLBT hydrodynamic model with quark coalescence~\cite{Zhao:2021vmu}. The left and right panels present the comparison for the 10--20\% and 40--50\% centrality intervals, respectively. The vertical error bars and the filled boxes represent statistical and systematic uncertainties of the data, respectively. The thickness of the model curves reflect the uncertainties of the hydrodynamic calculations.}
    \label{fig:v24Models}
\end{figure}

\begin{figure}[tb]
    \begin{center}
     \includegraphics[width = 0.49\textwidth,trim={0 0 1cm 1cm}, clip]{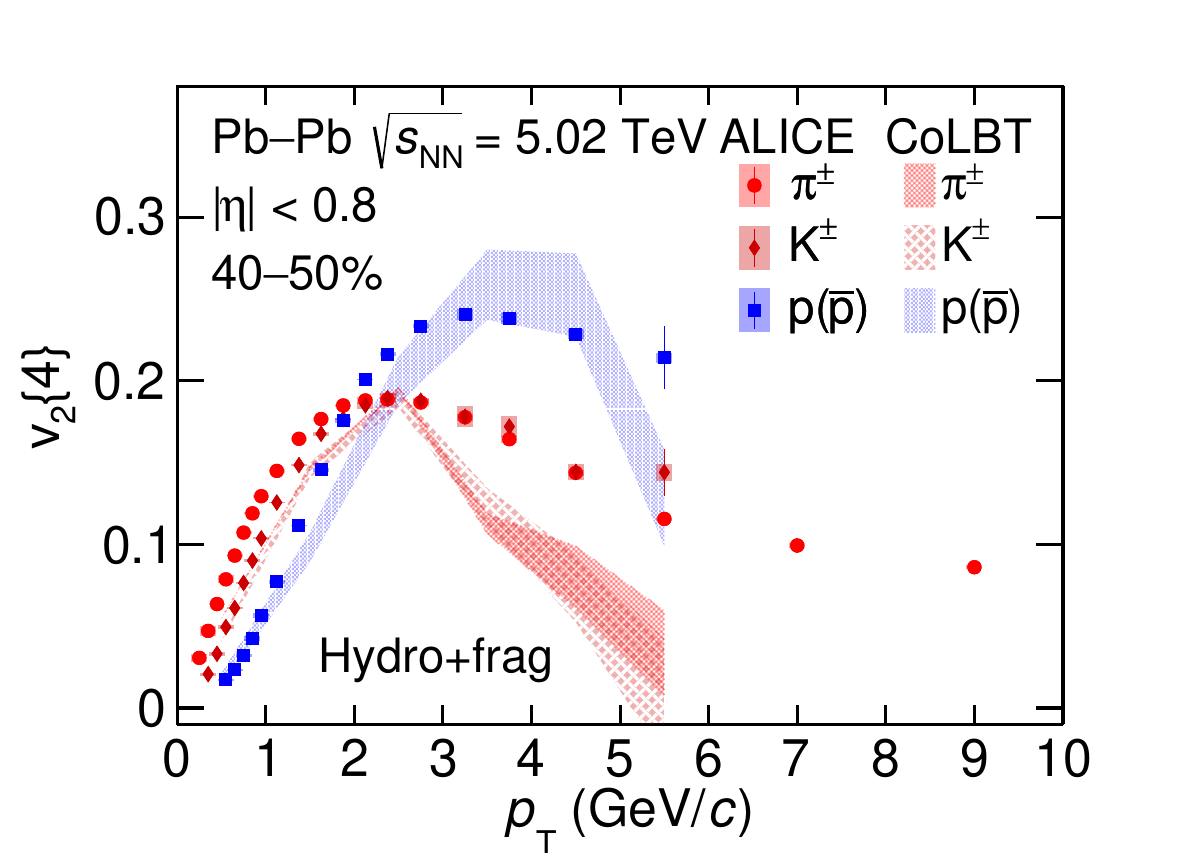}
  \end{center}
     \caption{The \pt-differential $v_2\{4\}$ for \pipm{}, \kapm{}, and p+\pbar{} measured in Pb--Pb collisions at $\sqrt{s_{\mathrm{NN}}} = 5.02$~TeV compared with expectations of the same quantity from the CoLBT hydrodynamic model without quark coalescence~\cite{Zhao:2021vmu} in 40--50\% centrality interval. The vertical error bars and the filled boxes represent statistical and systematic uncertainties of the data, respectively. The thickness of the model curves reflect the uncertainties of the hydrodynamic calculations.}
    \label{fig:v24Models-noCoal}
\end{figure}

\begin{figure}[tb]
    \begin{center}
    \includegraphics[width = 0.45\textwidth]{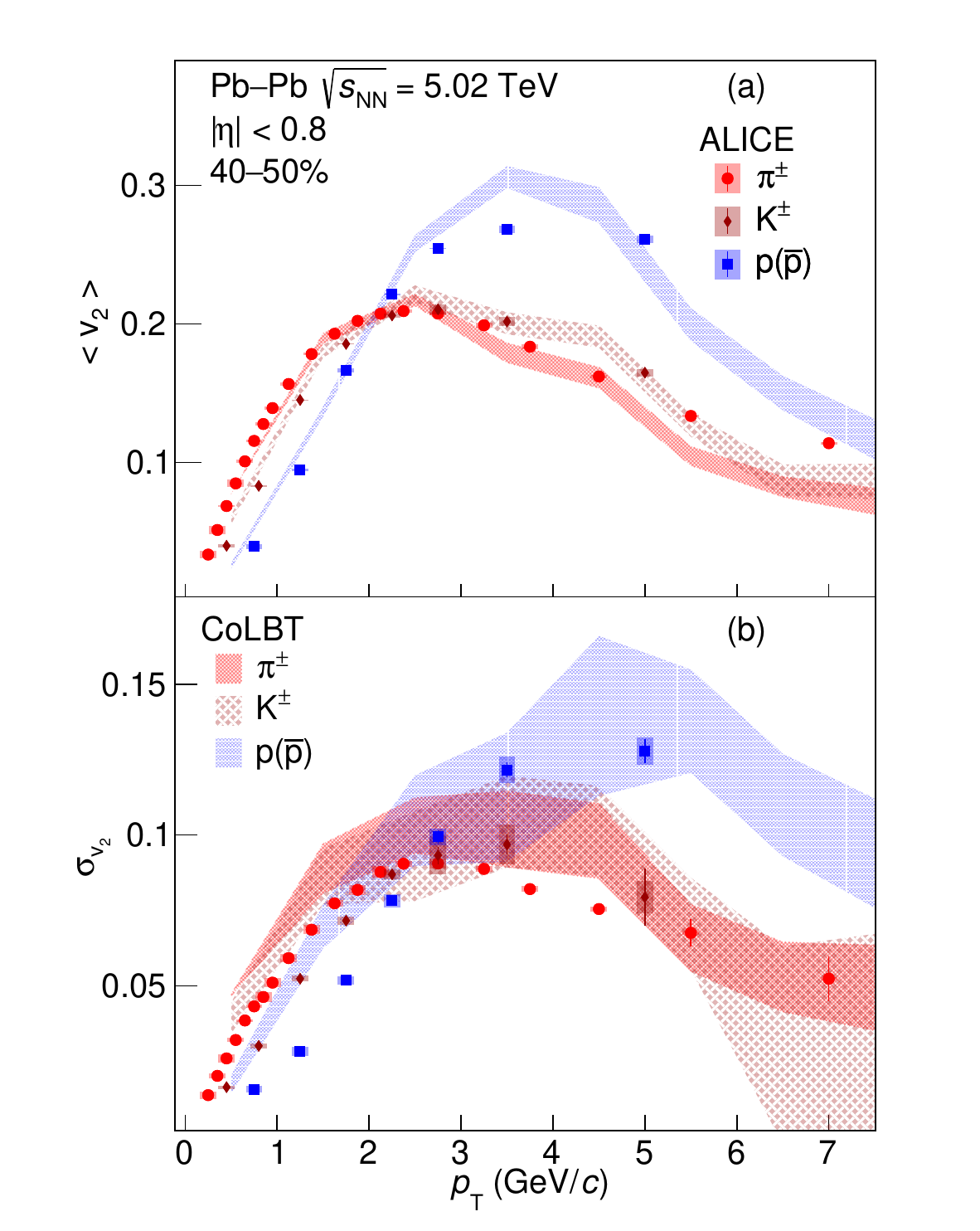}
    \includegraphics[width = 0.45\textwidth]{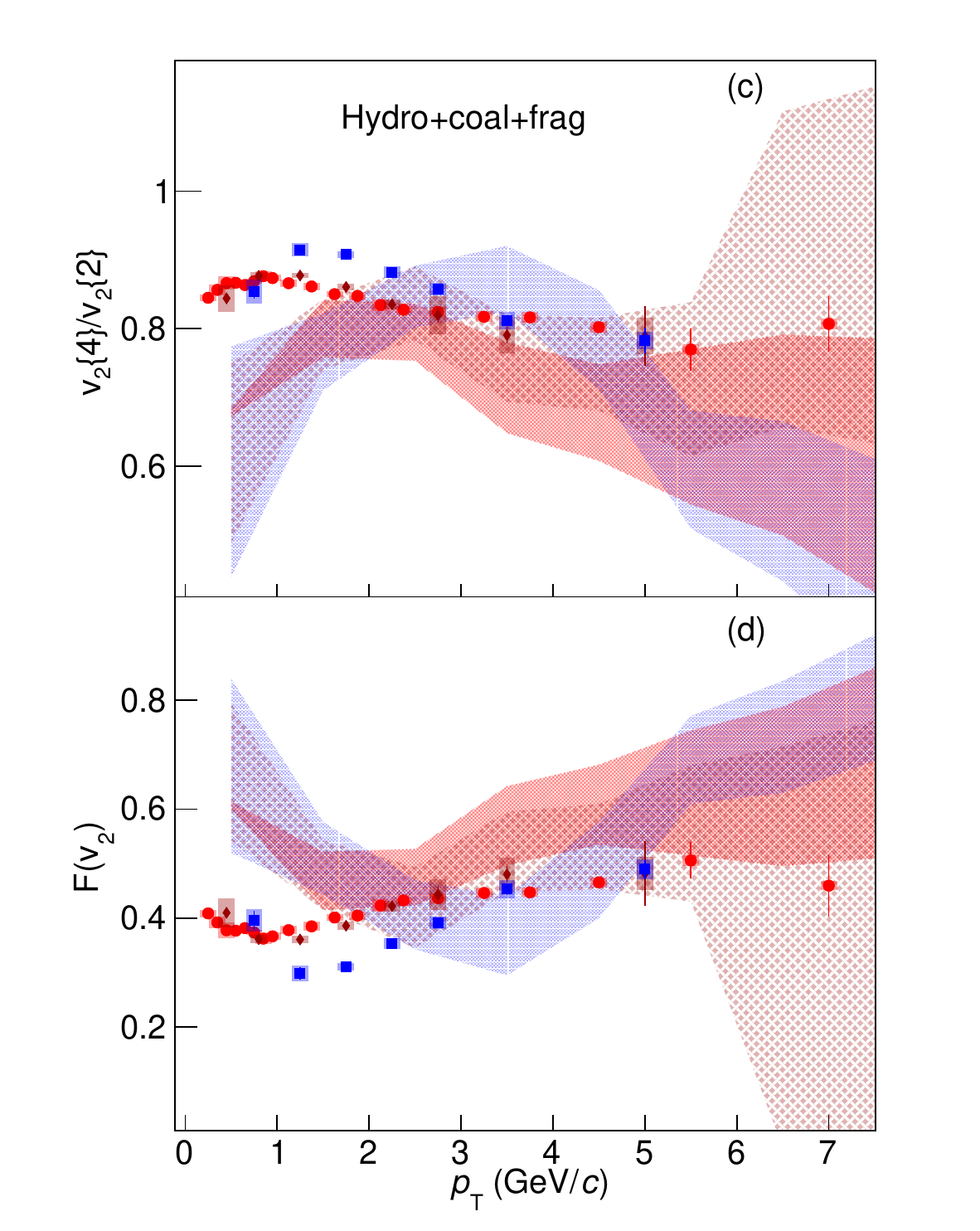}
    \end{center}
    \caption{The \pt-differential (a) $\langle v_2 \rangle$, (b)
    $\sigma_{v_2}$, (c) $v_2\{4\}/v_2\{2\}$, and (d) $F(v_2)$ 
    for \pipm{}, \kapm{}, and p+\pbar{} measured in one indicative 
    centrality interval (40--50\%) of Pb--Pb collisions at 
    $\sqrt{s_{\mathrm{NN}}} = 5.02$~TeV compared with expectations 
    of the same quantities from the CoLBT hydrodynamic 
    model~\cite{Zhao:2021vmu}. The vertical error bars and the filled boxes represent statistical and systematic uncertainties of the data, respectively. The thickness of the model curves reflect the uncertainties of the hydrodynamic calculations.}
    \label{fig:v24Models-2}
\end{figure}

The comparison of results from anisotropic flow studies with hydrodynamic calculations has been instrumental in constraining some of the basic transport coefficients of the QGP. However, such comparisons were limited until now to the low \pt{} region, i.e. in ranges where the mass ordering discussed in the previous section is prominent. One of the first attempts to provide a unified physics picture throughout the entire transverse momentum range for different particle species was presented recently in 
Ref.~\cite{Zhao:2021vmu}. In this article, the authors used the 
CoLBT hydrodynamic model~\cite{Chen:2020tbl} which allows for the simultaneous description of the evolution of parton showers and the bulk medium. The latter is prescribed by a \mbox{(3+1)-D} viscous hydrodynamic model that is initialized at $\tau_0 = 0.6$~fm/$c$ and uses a value of specific shear viscosity $\eta/s = 0.10$. The freeze-out temperature is set to $T_{\rm fo} = 150$~MeV, beyond which a hadronic after-burner describes the interactions between hadrons. The remaining parameters of the model were adjusted to reproduce the measured yields, \pt{} spectra, and integrated $v_{\rm n}$ of unidentified charged hadrons in Pb--Pb collisions. One of the important ingredients which is introduced in this model is the way hadrons emerge, with the typical hydrodynamic freeze-out at low \pt{} being complemented by a quark coalescence prescription at intermediate \pt{} and fragmentation at high \pt{}~\cite{Zhao:2021vmu}. 

Figure~\ref{fig:v24Models} presents the evolution of $v_2\{4\}$ as a function of \pt{} for \pipm{}, \kapm{}, and p+\pbar{} for two characteristic centrality intervals, 10--20\% (central) and 40--50\% (peripheral), in the left and right panels, respectively. The measurements are compared with the expectations for the same particle species from the CoLBT hydrodynamic model, represented by the shaded bands. It can be seen that the model describes the \pt{} dependence of $v_2\{4\}$ over the entire \pT~range. In particular, at low values of \pT~($<$~2--3~\GeVc{}) where the hydrodynamic expansion of the medium plays a dominant role, the model describes both the increase as a function of \pt{} and the mass ordering. The $v_{2}\{4\}$ reaches a peak value at around \pt{} $\approx$~3~\GeVc{} for pions and kaons and at \pt{} $\approx$~4~\GeVc{} for protons, before decreasing at high \pt{}. This can be naturally explained by the interplay between the hydrodynamical expansion, hadron production through quark coalescence, and jet fragmentation~\cite{Zhao:2021vmu}. 

Within the CoLBT model, the hydrodynamic contribution to $v_{2}$ is dominant for all particle species up to $\pt{} = 4$~\GeVc{}, whereas the jet fragmentation plays an increasingly important role for $\pt{} > 6$~\GeVc{}. In the intermediate \pT~region (4--6~\GeVc{}),  quark coalescence contributes in CoLBT to the development of the value of $v_{2}$, even though in the model this mechanism accounts for less than 25$\%$ of the total particle yield. This is because the value of $v_{2}$ from coalescence is significantly larger than the $v_{2}$ from fragmentation up to 6~\GeVc{}.
The additional mechanism of the coalescence prescription in the model is important to reproduce the experimental results quantitatively and to provide the proper connection between the low and high \pt{}~regions. In the former,  the mass ordering develops, while in the latter the fragmentation is the dominant particle production mechanism and no significant particle species dependence is observed. 

It is known that neither the hydrodynamic expansion nor the fragmentation alone leads to the precise NCQ scaling development. Such contributions in the final $v_{2}$ could consequently give a natural explanation for the significant deviation from a universal NCQ scaling observed in Fig.~\ref{fig:v24NCQ}. 
On the other hand, the model calculation with only contributions from hydrodynamic expansion and fragmentation but without the contribution from quark coalescence significantly underestimates $v_{2}\{4\}$ for \pt{} above 3 GeV/$c$ for \pipm{}, \kapm{}, shown in Fig.~\ref{fig:v24Models-noCoal}. Nevertheless, the crossing of $v_{2}$ of pions, kaons, and protons can develop according to CoLBT with a combination of the hydrodynamic expansion coupled only to jet fragmentation (for details refer to Fig. 4 in Ref.~\cite{Zhao:2021vmu}). This might challenge the prevailing idea discussed in the literature (see, e.g., Ref.~\cite{ALICE:2013snk}) that the crossing can be attributed to quark coalescence. It is important to note that the development of the crossing point in the absence of coalescence in CoLBT arises from a particle species-dependent \pT~value where fragmentation becomes dominant over hydrodynamics.

\begin{figure}[tb]
    \begin{center}
    \includegraphics[width = 0.45\textwidth]{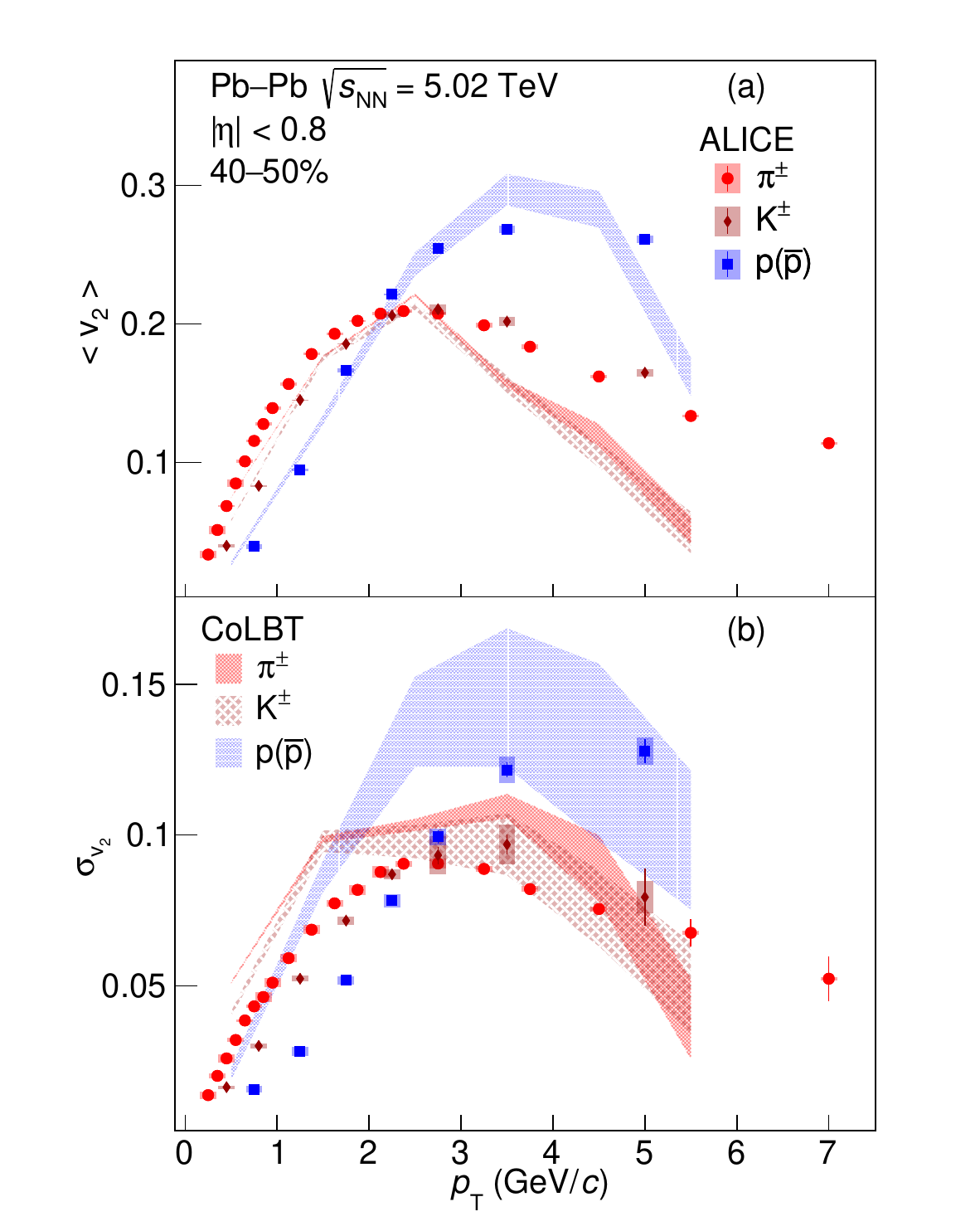}
    \includegraphics[width = 0.45\textwidth]{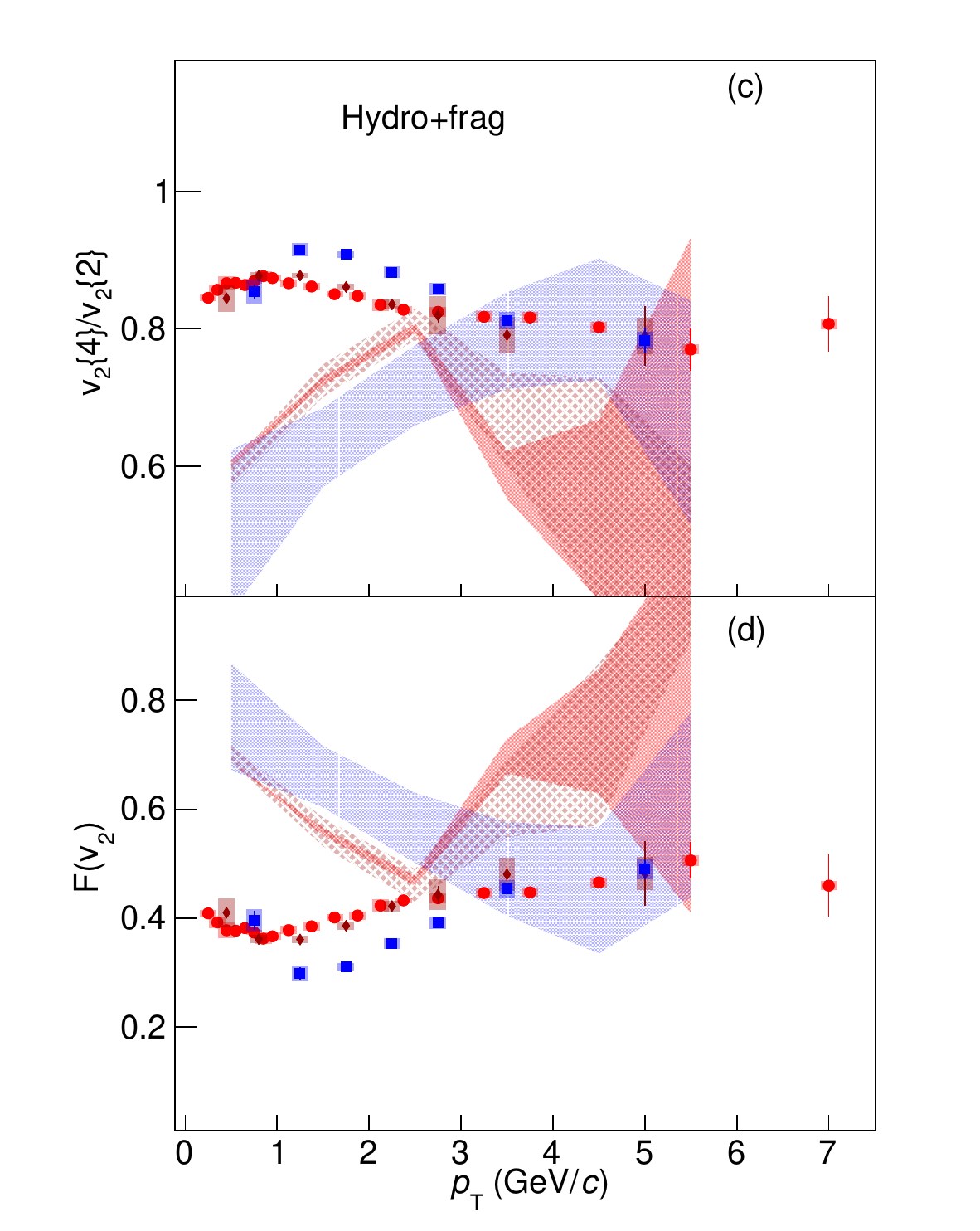}
    \end{center}
    \caption{The \pt-differential (a) $\langle v_2 \rangle$, (b)
    $\sigma_{v_2}$, (c) $v_2\{4\}/v_2\{2\}$, and (d) $F(v_2)$ 
    for \pipm{}, \kapm{}, and p+\pbar{} measured in one indicative 
    centrality interval (40--50\%) of Pb--Pb collisions at 
    $\sqrt{s_{\mathrm{NN}}} = 5.02$~TeV compared with expectations 
    of the same quantities from the CoLBT hydrodynamic 
    model without quark coalescence~\cite{Zhao:2021vmu}. The vertical error bars and the filled boxes represent statistical and systematic uncertainties of the data, respectively. The thickness of the model curves reflect the uncertainties of the hydrodynamic calculations.}
    \label{fig:v24Models-2-noCoal}
\end{figure}

To further investigate the coalescence contributions on flow fluctuations, Figs.~\ref{fig:v24Models-2} and~\ref{fig:v24Models-2-noCoal} present the comparison of the \pt-differential $\langle v_2 \rangle$ (panel a), $\sigma_{v_2}$ (panel b), $v_2\{4\}/v_2\{2\}$ (panel c), and $F(v_2)$ (panel d) for \pipm{}, \kapm{}, and p+\pbar{} with the calculation from CoLBT model with the combinations of hydrodynamics, quark coalescence, and jet fragmentation as well as with CoLBT model with only the combinations of hydrodynamics and jet fragmentation, respectively~\cite{Zhao:2021vmu}. The 40--50\% centrality interval was chosen as representative for these comparisons. The model without quark coalescence contribution describes qualitatively the features and the \pt{} dependence of the measurements, but significantly underestimates $\langle v_2 \rangle$ of pion and kaon for \pt{} above 3 GeV/$c$. This is very different from what has been observed in Figs.~\ref{fig:v24Models} and ~\ref{fig:v24Models-2}. Despite the sizable uncertainties of CoLBT calculations, the contribution from quark coalescence seems non-negligible for $\sigma_{v_2}$, $v_2\{4\}/v_2\{2\}$, and $F(v_2)$, when comparing the calculations of hydro+coal+frag (shown in Fig.~\ref{fig:v24Models-2}) and hydro+frag (shown in Fig.~\ref{fig:v24Models-2-noCoal}).

\section{Summary} 
\label{Sec:Summary}

In summary, the first measurement of \pt-differential elliptic flow using two- and four-particle cumulants for \pipm{}, \kapm{}, p+\pbar{}, \kzero{}, \lmb{}+\almb{}, $\phi$, \X{}+\Ix{}, and \Om{}+\Mo{} in Pb--Pb collisions at $\sqrt{s_{\rm NN}} = 5.02$~TeV is presented. The mean elliptic flow, elliptic flow fluctuations, and relative elliptic flow fluctuations are obtained for various particle species. Differences in the value of relative flow fluctuations for different particle species are observed, suggesting that final state hadronic interactions further modify the flow fluctuations. A distinct mass ordering is found in the 10--60\% centrality interval for $\pt < 3$~\GeVc{}, which arises from the interplay between the elliptic and radial flow. In the intermediate \pt{} range, the magnitude of $v_{2}\{4\}$, $\left< v_{2} \right>$, and $\sigma_{v_{2}}$ for baryons is larger than that for mesons by about 50\%. In addition, particles show an approximate constituent quark scaling. This scaling is tested for $v_{2}\{4\}$, which is expected to measure flow with little (or no) non-flow contamination. NCQ scaling describes the data no better than $\pm20\%$, an accuracy similar to what was reported for the $v_{2}$ using two-particle correlations. Furthermore, the relative flow fluctuation $F(v_2)$ for the identified hadrons shows an apparent splitting between baryons and mesons for centrality above 30\%, which suggests a significant role for final-state interactions in developing this observable.
Last but not least, CoLBT hydrodynamic calculations with the implementation of quark coalescence describe the measurements over a large \pt{} range, which confirms the relevance of the quark coalescence hadronization mechanism in the particle production in Pb--Pb collisions at the LHC.


\newenvironment{acknowledgement}{\relax}{\relax}
\begin{acknowledgement}
\section*{Acknowledgements}

The ALICE Collaboration would like to thank all its engineers and technicians for their invaluable contributions to the construction of the experiment and the CERN accelerator teams for the outstanding performance of the LHC complex.
The ALICE Collaboration gratefully acknowledges the resources and support provided by all Grid centres and the Worldwide LHC Computing Grid (WLCG) collaboration.
The ALICE Collaboration acknowledges the following funding agencies for their support in building and running the ALICE detector:
A. I. Alikhanyan National Science Laboratory (Yerevan Physics Institute) Foundation (ANSL), State Committee of Science and World Federation of Scientists (WFS), Armenia;
Austrian Academy of Sciences, Austrian Science Fund (FWF): [M 2467-N36] and Nationalstiftung f\"{u}r Forschung, Technologie und Entwicklung, Austria;
Ministry of Communications and High Technologies, National Nuclear Research Center, Azerbaijan;
Conselho Nacional de Desenvolvimento Cient\'{\i}fico e Tecnol\'{o}gico (CNPq), Financiadora de Estudos e Projetos (Finep), Funda\c{c}\~{a}o de Amparo \`{a} Pesquisa do Estado de S\~{a}o Paulo (FAPESP) and Universidade Federal do Rio Grande do Sul (UFRGS), Brazil;
Bulgarian Ministry of Education and Science, within the National Roadmap for Research Infrastructures 2020¿2027 (object CERN), Bulgaria;
Ministry of Education of China (MOEC) , Ministry of Science \& Technology of China (MSTC) and National Natural Science Foundation of China (NSFC), China;
Ministry of Science and Education and Croatian Science Foundation, Croatia;
Centro de Aplicaciones Tecnol\'{o}gicas y Desarrollo Nuclear (CEADEN), Cubaenerg\'{\i}a, Cuba;
Ministry of Education, Youth and Sports of the Czech Republic, Czech Republic;
The Danish Council for Independent Research | Natural Sciences, the VILLUM FONDEN and Danish National Research Foundation (DNRF), Denmark;
Helsinki Institute of Physics (HIP), Finland;
Commissariat \`{a} l'Energie Atomique (CEA) and Institut National de Physique Nucl\'{e}aire et de Physique des Particules (IN2P3) and Centre National de la Recherche Scientifique (CNRS), France;
Bundesministerium f\"{u}r Bildung und Forschung (BMBF) and GSI Helmholtzzentrum f\"{u}r Schwerionenforschung GmbH, Germany;
General Secretariat for Research and Technology, Ministry of Education, Research and Religions, Greece;
National Research, Development and Innovation Office, Hungary;
Department of Atomic Energy Government of India (DAE), Department of Science and Technology, Government of India (DST), University Grants Commission, Government of India (UGC) and Council of Scientific and Industrial Research (CSIR), India;
National Research and Innovation Agency - BRIN, Indonesia;
Istituto Nazionale di Fisica Nucleare (INFN), Italy;
Japanese Ministry of Education, Culture, Sports, Science and Technology (MEXT) and Japan Society for the Promotion of Science (JSPS) KAKENHI, Japan;
Consejo Nacional de Ciencia (CONACYT) y Tecnolog\'{i}a, through Fondo de Cooperaci\'{o}n Internacional en Ciencia y Tecnolog\'{i}a (FONCICYT) and Direcci\'{o}n General de Asuntos del Personal Academico (DGAPA), Mexico;
Nederlandse Organisatie voor Wetenschappelijk Onderzoek (NWO), Netherlands;
The Research Council of Norway, Norway;
Commission on Science and Technology for Sustainable Development in the South (COMSATS), Pakistan;
Pontificia Universidad Cat\'{o}lica del Per\'{u}, Peru;
Ministry of Education and Science, National Science Centre and WUT ID-UB, Poland;
Korea Institute of Science and Technology Information and National Research Foundation of Korea (NRF), Republic of Korea;
Ministry of Education and Scientific Research, Institute of Atomic Physics, Ministry of Research and Innovation and Institute of Atomic Physics and University Politehnica of Bucharest, Romania;
Ministry of Education, Science, Research and Sport of the Slovak Republic, Slovakia;
National Research Foundation of South Africa, South Africa;
Swedish Research Council (VR) and Knut \& Alice Wallenberg Foundation (KAW), Sweden;
European Organization for Nuclear Research, Switzerland;
Suranaree University of Technology (SUT), National Science and Technology Development Agency (NSTDA), Thailand Science Research and Innovation (TSRI) and National Science, Research and Innovation Fund (NSRF), Thailand;
Turkish Energy, Nuclear and Mineral Research Agency (TENMAK), Turkey;
National Academy of  Sciences of Ukraine, Ukraine;
Science and Technology Facilities Council (STFC), United Kingdom;
National Science Foundation of the United States of America (NSF) and United States Department of Energy, Office of Nuclear Physics (DOE NP), United States of America.
In addition, individual groups or members have received support from:
Marie Sk\l{}odowska Curie, European Research Council, Strong 2020 - Horizon 2020 (grant nos. 950692, 824093, 896850), European Union;
Academy of Finland (Center of Excellence in Quark Matter) (grant nos. 346327, 346328), Finland;
Programa de Apoyos para la Superaci\'{o}n del Personal Acad\'{e}mico, UNAM, Mexico.

\end{acknowledgement}

\bibliographystyle{utphys}   
\bibliography{bibliography}

\providecommand{\href}[2]{#2}\begingroup\raggedright\begin{thebibliography}{10}

\bibitem{Shuryak:1978ij}
E.~V. Shuryak, ``{Quark-Gluon Plasma and Hadronic Production of Leptons,
  Photons and Pions}'',
  \href{http://dx.doi.org/10.1016/0370-2693(78)90370-2}{{\em Phys. Lett.}
  {\bfseries B78} (1978) 150}.
[Yad. Fiz.28,796(1978)].

\bibitem{Shuryak:1980tp}
E.~V. Shuryak, ``{Quantum Chromodynamics and the Theory of Superdense
  Matter}'',
\href{http://dx.doi.org/10.1016/0370-1573(80)90105-2}{{\em Phys. Rept.}
  {\bfseries 61} (1980) 71--158}.

\bibitem{Ollitrault:1992bk}
J.-Y. Ollitrault, ``{Anisotropy as a signature of transverse collective
  flow}'',
\href{http://dx.doi.org/10.1103/PhysRevD.46.229}{{\em Phys. Rev.} {\bfseries
  D46} (1992) 229--245}.

\bibitem{Voloshin:2008dg}
S.~A. Voloshin, A.~M. Poskanzer, and R.~Snellings, ``{Collective phenomena in
  non-central nuclear collisions}'',
  \href{http://dx.doi.org/10.1007/978-3-642-01539-7_10}{{\em Landolt-Bornstein}
  {\bfseries 23} (2010) 293--333},
  \href{http://arxiv.org/abs/0809.2949}{{\ttfamily arXiv:0809.2949 [nucl-ex]}}.

\bibitem{Shuryak:1984nq}
E.~V. Shuryak, ``{Theory and phenomenology of the QCD vacuum}'',
\href{http://dx.doi.org/10.1016/0370-1573(84)90037-1}{{\em Phys. Rept.}
  {\bfseries 115} (1984) 151}.

\bibitem{Cleymans:1985wb}
J.~Cleymans, R.~V. Gavai, and E.~Suhonen, ``{Quarks and Gluons at High
  Temperatures and Densities}'',
\href{http://dx.doi.org/10.1016/0370-1573(86)90169-9}{{\em Phys. Rept.}
  {\bfseries 130} (1986) 217}.

\bibitem{Bass:1998vz}
S.~A. Bass, M.~Gyulassy, H.~Stoecker, and W.~Greiner, ``{Signatures of quark
  gluon plasma formation in high-energy heavy ion collisions: A Critical
  review}'', \href{http://dx.doi.org/10.1088/0954-3899/25/3/013}{{\em J. Phys.}
  {\bfseries G25} (1999) R1--R57},
\href{http://arxiv.org/abs/hep-ph/9810281}{{\ttfamily arXiv:hep-ph/9810281
  [hep-ph]}}.

\bibitem{Voloshin:1994mz}
S.~Voloshin and Y.~Zhang, ``{Flow study in relativistic nuclear collisions by
  Fourier expansion of Azimuthal particle distributions}'',
  \href{http://dx.doi.org/10.1007/s002880050141}{{\em Z. Phys.} {\bfseries C70}
  (1996) 665--672},
\href{http://arxiv.org/abs/hep-ph/9407282}{{\ttfamily arXiv:hep-ph/9407282
  [hep-ph]}}.

\bibitem{Poskanzer:1998yz}
A.~M. Poskanzer and S.~A. Voloshin, ``{Methods for analyzing anisotropic flow
  in relativistic nuclear collisions}'',
  \href{http://dx.doi.org/10.1103/PhysRevC.58.1671}{{\em Phys. Rev.} {\bfseries
  C58} (1998) 1671--1678},
\href{http://arxiv.org/abs/nucl-ex/9805001}{{\ttfamily arXiv:nucl-ex/9805001
  [nucl-ex]}}.

\bibitem{Alver:2010gr}
B.~Alver and G.~Roland, ``{Collision geometry fluctuations and triangular flow
  in heavy-ion collisions}'',
  \href{http://dx.doi.org/10.1103/PhysRevC.82.039903,
  10.1103/PhysRevC.81.054905}{{\em Phys. Rev.} {\bfseries C81} (2010) 054905},
  \href{http://arxiv.org/abs/1003.0194}{{\ttfamily arXiv:1003.0194 [nucl-th]}}.
[Erratum: Phys. Rev.C82,039903(2010)].

\bibitem{Alver:2010dn}
B.~Alver, C.~Gombeaud, M.~Luzum, and J.-Y. Ollitrault, ``{Triangular flow in
  hydrodynamics and transport theory}'',
  \href{http://dx.doi.org/10.1103/PhysRevC.82.034913}{{\em Phys. Rev.}
  {\bfseries C82} (2010) 034913},
\href{http://arxiv.org/abs/1007.5469}{{\ttfamily arXiv:1007.5469 [nucl-th]}}.

\bibitem{Arsene:2004fa}
{\bfseries BRAHMS} Collaboration, I.~Arsene {\em et~al.}, ``{Quark gluon plasma
  and color glass condensate at RHIC? The Perspective from the BRAHMS
  experiment}'', \href{http://dx.doi.org/10.1016/j.nuclphysa.2005.02.130}{{\em
  Nucl. Phys. A} {\bfseries 757} (2005) 1--27},
  \href{http://arxiv.org/abs/nucl-ex/0410020}{{\ttfamily
  arXiv:nucl-ex/0410020}}.

\bibitem{Adcox:2004mh}
{\bfseries PHENIX} Collaboration, K.~Adcox {\em et~al.}, ``{Formation of dense
  partonic matter in relativistic nucleus-nucleus collisions at RHIC:
  Experimental evaluation by the PHENIX collaboration}'',
  \href{http://dx.doi.org/10.1016/j.nuclphysa.2005.03.086}{{\em Nucl. Phys. A}
  {\bfseries 757} (2005) 184--283},
  \href{http://arxiv.org/abs/nucl-ex/0410003}{{\ttfamily
  arXiv:nucl-ex/0410003}}.

\bibitem{Back:2004je}
{\bfseries PHOBOS} Collaboration, B.~B. Back {\em et~al.}, ``{The PHOBOS
  perspective on discoveries at RHIC}'',
  \href{http://dx.doi.org/10.1016/j.nuclphysa.2005.03.084}{{\em Nucl. Phys. A}
  {\bfseries 757} (2005) 28--101},
  \href{http://arxiv.org/abs/nucl-ex/0410022}{{\ttfamily
  arXiv:nucl-ex/0410022}}.

\bibitem{Adams:2005dq}
{\bfseries STAR} Collaboration, J.~Adams {\em et~al.}, ``{Experimental and
  theoretical challenges in the search for the quark gluon plasma: The STAR
  Collaboration's critical assessment of the evidence from RHIC collisions}'',
  \href{http://dx.doi.org/10.1016/j.nuclphysa.2005.03.085}{{\em Nucl. Phys. A}
  {\bfseries 757} (2005) 102--183},
  \href{http://arxiv.org/abs/nucl-ex/0501009}{{\ttfamily
  arXiv:nucl-ex/0501009}}.

\bibitem{ALICE:2011ab}
{\bfseries ALICE} Collaboration, K.~Aamodt {\em et~al.}, ``{Higher harmonic
  anisotropic flow measurements of charged particles in Pb-Pb collisions at
  $\sqrt{s_{NN}}$ = 2.76 TeV}'',
  \href{http://dx.doi.org/10.1103/PhysRevLett.107.032301}{{\em Phys. Rev.
  Lett.} {\bfseries 107} (2011) 032301},
\href{http://arxiv.org/abs/1105.3865}{{\ttfamily arXiv:1105.3865 [nucl-ex]}}.

\bibitem{ATLAS:2012at}
{\bfseries ATLAS} Collaboration, G.~Aad {\em et~al.}, ``{Measurement of the
  azimuthal anisotropy for charged particle production in $\sqrt{s_{NN}}=2.76$
  TeV lead-lead collisions with the ATLAS detector}'',
  \href{http://dx.doi.org/10.1103/PhysRevC.86.014907}{{\em Phys. Rev.}
  {\bfseries C86} (2012) 014907},
\href{http://arxiv.org/abs/1203.3087}{{\ttfamily arXiv:1203.3087 [hep-ex]}}.

\bibitem{Chatrchyan:2013kba}
{\bfseries CMS} Collaboration, S.~Chatrchyan {\em et~al.}, ``{Measurement of
  higher-order harmonic azimuthal anisotropy in PbPb collisions at
  $\sqrt{s_{NN}}$ = 2.76 TeV}'',
  \href{http://dx.doi.org/10.1103/PhysRevC.89.044906}{{\em Phys. Rev.}
  {\bfseries C89} (2014) 044906},
\href{http://arxiv.org/abs/1310.8651}{{\ttfamily arXiv:1310.8651 [nucl-ex]}}.

\bibitem{Gardim:2011xv}
F.~G. Gardim, F.~Grassi, M.~Luzum, and J.-Y. Ollitrault, ``{Mapping the
  hydrodynamic response to the initial geometry in heavy-ion collisions}'',
  \href{http://dx.doi.org/10.1103/PhysRevC.85.024908}{{\em Phys. Rev. C}
  {\bfseries 85} (2012) 024908},
  \href{http://arxiv.org/abs/1111.6538}{{\ttfamily arXiv:1111.6538 [nucl-th]}}.

\bibitem{CMS:2015xmx}
{\bfseries CMS} Collaboration, V.~Khachatryan {\em et~al.}, ``{Evidence for
  transverse momentum and pseudorapidity dependent event plane fluctuations in
  PbPb and pPb collisions}'',
  \href{http://dx.doi.org/10.1103/PhysRevC.92.034911}{{\em Phys. Rev. C}
  {\bfseries 92} (2015) 034911},
  \href{http://arxiv.org/abs/1503.01692}{{\ttfamily arXiv:1503.01692
  [nucl-ex]}}.

\bibitem{ALICE:2017lyf}
{\bfseries ALICE} Collaboration, S.~Acharya {\em et~al.}, ``{Searches for
  transverse momentum dependent flow vector fluctuations in Pb-Pb and p-Pb
  collisions at the LHC}'',
  \href{http://dx.doi.org/10.1007/JHEP09(2017)032}{{\em JHEP} {\bfseries 09}
  (2017) 032}, \href{http://arxiv.org/abs/1707.05690}{{\ttfamily
  arXiv:1707.05690 [nucl-ex]}}.

\bibitem{ATLAS:2017rij}
{\bfseries ATLAS} Collaboration, M.~Aaboud {\em et~al.}, ``{Measurement of
  longitudinal flow decorrelations in Pb+Pb collisions at $\sqrt{s_{\text
  {NN}}}=2.76$ and 5.02 TeV with the ATLAS detector}'',
  \href{http://dx.doi.org/10.1140/epjc/s10052-018-5605-7}{{\em Eur. Phys. J. C}
  {\bfseries 78} (2018) 142}, \href{http://arxiv.org/abs/1709.02301}{{\ttfamily
  arXiv:1709.02301 [nucl-ex]}}.

\bibitem{Acharya:2018lmh}
{\bfseries ALICE} Collaboration, S.~Acharya {\em et~al.}, ``{Energy dependence
  and fluctuations of anisotropic flow in Pb-Pb collisions at $
  \sqrt{s_{\mathrm{NN}}}=5.02 $ and 2.76 TeV}'',
  \href{http://dx.doi.org/10.1007/JHEP07(2018)103}{{\em JHEP} {\bfseries 07}
  (2018) 103}, \href{http://arxiv.org/abs/1804.02944}{{\ttfamily
  arXiv:1804.02944 [nucl-ex]}}.

\bibitem{Sirunyan:2017fts}
{\bfseries CMS} Collaboration, A.~M. Sirunyan {\em et~al.}, ``{Non-Gaussian
  elliptic-flow fluctuations in PbPb collisions at
  $\sqrt{\smash[b]{s_{_\text{NN}}}} = 5.02$ TeV}'',
  \href{http://dx.doi.org/10.1016/j.physletb.2018.11.063}{{\em Phys. Lett. B}
  {\bfseries 789} (2019) 643--665},
  \href{http://arxiv.org/abs/1711.05594}{{\ttfamily arXiv:1711.05594
  [nucl-ex]}}.

\bibitem{ALICE:2012vgf}
{\bfseries ALICE} Collaboration, B.~Abelev {\em et~al.}, ``{Anisotropic flow of
  charged hadrons, pions and (anti-)protons measured at high transverse
  momentum in Pb-Pb collisions at $\sqrt{s_{NN}}$=2.76 TeV}'',
  \href{http://dx.doi.org/10.1016/j.physletb.2012.12.066}{{\em Phys. Lett. B}
  {\bfseries 719} (2013) 18--28},
  \href{http://arxiv.org/abs/1205.5761}{{\ttfamily arXiv:1205.5761 [nucl-ex]}}.

\bibitem{Abelev:2014pua}
{\bfseries ALICE} Collaboration, B.~Abelev {\em et~al.}, ``{Elliptic flow of
  identified hadrons in Pb-Pb collisions at $ \sqrt{s_{\mathrm{NN}}}=2.76 $
  TeV}'', \href{http://dx.doi.org/10.1007/JHEP06(2015)190}{{\em JHEP}
  {\bfseries 06} (2015) 190}, \href{http://arxiv.org/abs/1405.4632}{{\ttfamily
  arXiv:1405.4632 [nucl-ex]}}.

\bibitem{ALICE:2016cti}
{\bfseries ALICE} Collaboration, J.~Adam {\em et~al.}, ``{Higher harmonic flow
  coefficients of identified hadrons in Pb-Pb collisions at $\sqrt{s_{\rm NN}}$
  = 2.76 TeV}'', \href{http://dx.doi.org/10.1007/JHEP09(2016)164}{{\em JHEP}
  {\bfseries 09} (2016) 164}, \href{http://arxiv.org/abs/1606.06057}{{\ttfamily
  arXiv:1606.06057 [nucl-ex]}}.

\bibitem{ALICE:2018yph}
{\bfseries ALICE} Collaboration, S.~Acharya {\em et~al.}, ``{Anisotropic flow
  of identified particles in Pb-Pb collisions at $ {\sqrt{s}}_{\mathrm{NN}}$ =
  5.02 TeV}'', \href{http://dx.doi.org/10.1007/JHEP09(2018)006}{{\em JHEP}
  {\bfseries 09} (2018) 006}, \href{http://arxiv.org/abs/1805.04390}{{\ttfamily
  arXiv:1805.04390 [nucl-ex]}}.

\bibitem{Kovtun:2004de}
P.~Kovtun, D.~T. Son, and A.~O. Starinets, ``{Viscosity in strongly interacting
  quantum field theories from black hole physics}'',
  \href{http://dx.doi.org/10.1103/PhysRevLett.94.111601}{{\em Phys. Rev. Lett.}
  {\bfseries 94} (2005) 111601},
\href{http://arxiv.org/abs/hep-th/0405231}{{\ttfamily arXiv:hep-th/0405231
  [hep-th]}}.

\bibitem{Huovinen:2001cy}
P.~Huovinen, P.~F. Kolb, U.~W. Heinz, P.~V. Ruuskanen, and S.~A. Voloshin,
  ``{Radial and elliptic flow at RHIC: Further predictions}'',
  \href{http://dx.doi.org/10.1016/S0370-2693(01)00219-2}{{\em Phys. Lett. B}
  {\bfseries 503} (2001) 58--64},
  \href{http://arxiv.org/abs/hep-ph/0101136}{{\ttfamily arXiv:hep-ph/0101136}}.

\bibitem{Voloshin:2002wa}
S.~A. Voloshin, ``{Anisotropic flow}'',
  \href{http://dx.doi.org/10.1016/S0375-9474(02)01450-1}{{\em Nucl. Phys.}
  {\bfseries A715} (2003) 379--388},
\href{http://arxiv.org/abs/nucl-ex/0210014}{{\ttfamily arXiv:nucl-ex/0210014
  [nucl-ex]}}.

\bibitem{Molnar:2003ff}
D.~Molnar and S.~A. Voloshin, ``{Elliptic flow at large transverse momenta from
  quark coalescence}'',
  \href{http://dx.doi.org/10.1103/PhysRevLett.91.092301}{{\em Phys. Rev. Lett.}
  {\bfseries 91} (2003) 092301},
\href{http://arxiv.org/abs/nucl-th/0302014}{{\ttfamily arXiv:nucl-th/0302014
  [nucl-th]}}.

\bibitem{ALICE:2019xkq}
{\bfseries ALICE} Collaboration, S.~Acharya {\em et~al.}, ``{Non-linear flow
  modes of identified particles in Pb-Pb collisions at $ \sqrt{s_{\mathrm{NN}}}
  $ = 5.02 TeV}'', \href{http://dx.doi.org/10.1007/JHEP06(2020)147}{{\em JHEP}
  {\bfseries 06} (2020) 147}, \href{http://arxiv.org/abs/1912.00740}{{\ttfamily
  arXiv:1912.00740 [nucl-ex]}}.

\bibitem{Bhalerao:2014xra}
R.~S. Bhalerao, J.-Y. Ollitrault, and S.~Pal, ``{Characterizing flow
  fluctuations with moments}'',
  \href{http://dx.doi.org/10.1016/j.physletb.2015.01.019}{{\em Phys. Lett. B}
  {\bfseries 742} (2015) 94--98},
  \href{http://arxiv.org/abs/1411.5160}{{\ttfamily arXiv:1411.5160 [nucl-th]}}.

\bibitem{Yan:2015jma}
L.~Yan and J.-Y. Ollitrault, ``{$\nu_4, \nu_5, \nu_6, \nu_7$: nonlinear
  hydrodynamic response versus LHC data}'',
  \href{http://dx.doi.org/10.1016/j.physletb.2015.03.040}{{\em Phys. Lett. B}
  {\bfseries 744} (2015) 82--87},
  \href{http://arxiv.org/abs/1502.02502}{{\ttfamily arXiv:1502.02502
  [nucl-th]}}.

\bibitem{ALICE:2017fcd}
{\bfseries ALICE} Collaboration, S.~Acharya {\em et~al.}, ``{Linear and
  non-linear flow modes in Pb-Pb collisions at $\sqrt{s_{\rm NN}} =$ 2.76
  TeV}'', \href{http://dx.doi.org/10.1016/j.physletb.2017.07.060}{{\em Phys.
  Lett. B} {\bfseries 773} (2017) 68--80},
  \href{http://arxiv.org/abs/1705.04377}{{\ttfamily arXiv:1705.04377
  [nucl-ex]}}.

\bibitem{ALICE:2020sup}
{\bfseries ALICE} Collaboration, S.~Acharya {\em et~al.}, ``{Higher harmonic
  non-linear flow modes of charged hadrons in Pb-Pb collisions at
  $\sqrt{s_{\rm{NN}}}$ = 5.02 TeV}'',
  \href{http://dx.doi.org/10.1007/JHEP05(2020)085}{{\em JHEP} {\bfseries 05}
  (2020) 085}, \href{http://arxiv.org/abs/2002.00633}{{\ttfamily
  arXiv:2002.00633 [nucl-ex]}}.

\bibitem{Bilandzic:2013kga}
A.~Bilandzic, C.~H. Christensen, K.~Gulbrandsen, A.~Hansen, and Y.~Zhou,
  ``{Generic framework for anisotropic flow analyses with multiparticle
  azimuthal correlations}'',
  \href{http://dx.doi.org/10.1103/PhysRevC.89.064904}{{\em Phys. Rev. C}
  {\bfseries 89} (2014) 064904},
  \href{http://arxiv.org/abs/1312.3572}{{\ttfamily arXiv:1312.3572 [nucl-ex]}}.

\bibitem{Huo:2017nms}
P.~Huo, K.~Gajdo\v{s}ov\'a, J.~Jia, and Y.~Zhou, ``{Importance of non-flow in
  mixed-harmonic multi-particle correlations in small collision systems}'',
  \href{http://dx.doi.org/10.1016/j.physletb.2017.12.035}{{\em Phys. Lett. B}
  {\bfseries 777} (2018) 201--206},
  \href{http://arxiv.org/abs/1710.07567}{{\ttfamily arXiv:1710.07567
  [nucl-ex]}}.

\bibitem{Moravcova:2020wnf}
Z.~Moravcova, K.~Gulbrandsen, and Y.~Zhou, ``{Generic algorithm for
  multiparticle cumulants of azimuthal correlations in high energy nucleus
  collisions}'', \href{http://dx.doi.org/10.1103/PhysRevC.103.024913}{{\em
  Phys. Rev. C} {\bfseries 103} (2021) 024913},
  \href{http://arxiv.org/abs/2005.07974}{{\ttfamily arXiv:2005.07974
  [nucl-th]}}.

\bibitem{Aamodt:2008zz}
{\bfseries ALICE} Collaboration, K.~Aamodt {\em et~al.}, ``{The ALICE
  experiment at the CERN LHC}'',
\href{http://dx.doi.org/10.1088/1748-0221/3/08/S08002}{{\em JINST} {\bfseries
  3} (2008) S08002}.

\bibitem{Abelev:2014ffa}
{\bfseries ALICE} Collaboration, B.~Abelev {\em et~al.}, ``{Performance of the
  ALICE Experiment at the CERN LHC}'',
  \href{http://dx.doi.org/10.1142/S0217751X14300440}{{\em Int. J. Mod. Phys.}
  {\bfseries A29} (2014) 1430044},
\href{http://arxiv.org/abs/1402.4476}{{\ttfamily arXiv:1402.4476 [nucl-ex]}}.

\bibitem{Alme:2010ke}
J.~Alme {\em et~al.}, ``{The ALICE TPC, a large 3-dimensional tracking device
  with fast readout for ultra-high multiplicity events}'',
  \href{http://dx.doi.org/10.1016/j.nima.2010.04.042}{{\em Nucl. Instrum.
  Meth.} {\bfseries A622} (2010) 316--367},
\href{http://arxiv.org/abs/1001.1950}{{\ttfamily arXiv:1001.1950
  [physics.ins-det]}}.

\bibitem{Akindinov:2013tea}
A.~Akindinov {\em et~al.}, ``{Performance of the ALICE Time-Of-Flight detector
  at the LHC}'', \href{http://dx.doi.org/10.1140/epjp/i2013-13044-x}{{\em Eur.
  Phys. J. Plus} {\bfseries 128} (2013) 44}.

\bibitem{Bondila:2005xy}
M.~Bondila {\em et~al.}, ``{ALICE T0 detector}'',
  \href{http://dx.doi.org/10.1109/TNS.2005.856900}{{\em IEEE Trans. Nucl. Sci.}
  {\bfseries 52} (2005) 1705--1711}.

\bibitem{Abbas:2013taa}
{\bfseries ALICE} Collaboration, E.~Abbas {\em et~al.}, ``{Performance of the
  ALICE VZERO system}'',
  \href{http://dx.doi.org/10.1088/1748-0221/8/10/P10016}{{\em JINST} {\bfseries
  8} (2013) P10016},
\href{http://arxiv.org/abs/1306.3130}{{\ttfamily arXiv:1306.3130 [nucl-ex]}}.

\bibitem{ALICE:2013hur}
{\bfseries ALICE} Collaboration, B.~Abelev {\em et~al.}, ``{Centrality
  determination of Pb-Pb collisions at \mbox{$\sqrt{s_{NN}}$ = 2.76 TeV} with
  ALICE}'', \href{http://dx.doi.org/10.1103/PhysRevC.88.044909}{{\em Phys. Rev.
  C} {\bfseries 88} (2013) 044909},
  \href{http://arxiv.org/abs/1301.4361}{{\ttfamily arXiv:1301.4361 [nucl-ex]}}.

\bibitem{ALICE:BayesPID}
{\bfseries ALICE} Collaboration, J.~Adam and others., ``{Particle
  identification in ALICE: a Bayesian approach}'',
  \href{http://dx.doi.org/10.1140/epjp/i2016-16168-5}{{\em Eur. Phys. J. Plus}
  {\bfseries 131} (2016) }, \href{http://arxiv.org/abs/1602.01392}{{\ttfamily
  arXiv:1602.01392}}.

\bibitem{Zyla:2020zbs}
{\bfseries Particle Data Group} Collaboration, P.~Zyla {\em et~al.}, ``{Review
  of Particle Physics}'', \href{http://dx.doi.org/10.1093/ptep/ptaa104}{{\em
  PTEP} {\bfseries 2020} (2020) 083C01}.

\bibitem{armpod}
J.~Podolanski and R.~Armenteros, ``Iii. analysis of v-events'',
  \href{http://dx.doi.org/10.1080/14786440108520416}{{\em The London,
  Edinburgh, and Dublin Philosophical Magazine and Journal of Science}
  {\bfseries 45} (1954) 13--30}.

\bibitem{Borghini:2000sa}
N.~Borghini, P.~M. Dinh, and J.-Y. Ollitrault, ``{A New method for measuring
  azimuthal distributions in nucleus-nucleus collisions}'',
  \href{http://dx.doi.org/10.1103/PhysRevC.63.054906}{{\em Phys. Rev. C}
  {\bfseries 63} (2001) 054906},
  \href{http://arxiv.org/abs/nucl-th/0007063}{{\ttfamily
  arXiv:nucl-th/0007063}}.

\bibitem{Borghini:2001vi}
N.~Borghini, P.~M. Dinh, and J.-Y. Ollitrault, ``{Flow analysis from
  multiparticle azimuthal correlations}'',
  \href{http://dx.doi.org/10.1103/PhysRevC.64.054901}{{\em Phys. Rev. C}
  {\bfseries 64} (2001) 054901},
  \href{http://arxiv.org/abs/nucl-th/0105040}{{\ttfamily
  arXiv:nucl-th/0105040}}.

\bibitem{ATLAS:2013xzf}
{\bfseries ATLAS} Collaboration, G.~Aad {\em et~al.}, ``{Measurement of the
  distributions of event-by-event flow harmonics in lead-lead collisions at =
  2.76 TeV with the ATLAS detector at the LHC}'',
  \href{http://dx.doi.org/10.1007/JHEP11(2013)183}{{\em JHEP} {\bfseries 11}
  (2013) 183}, \href{http://arxiv.org/abs/1305.2942}{{\ttfamily arXiv:1305.2942
  [hep-ex]}}.

\bibitem{ALICE:2018rtz}
{\bfseries ALICE} Collaboration, S.~Acharya {\em et~al.}, ``{Energy dependence
  and fluctuations of anisotropic flow in Pb-Pb collisions at $
  \sqrt{s_{\mathrm{NN}}}=5.02 $ and 2.76 TeV}'',
  \href{http://dx.doi.org/10.1007/JHEP07(2018)103}{{\em JHEP} {\bfseries 07}
  (2018) 103}, \href{http://arxiv.org/abs/1804.02944}{{\ttfamily
  arXiv:1804.02944 [nucl-ex]}}.

\bibitem{CMS:2017glf}
{\bfseries CMS} Collaboration, A.~M. Sirunyan {\em et~al.}, ``{Non-Gaussian
  elliptic-flow fluctuations in PbPb collisions at
  $\sqrt{\smash[b]{s_{_\text{NN}}}} = 5.02$ TeV}'',
  \href{http://dx.doi.org/10.1016/j.physletb.2018.11.063}{{\em Phys. Lett. B}
  {\bfseries 789} (2019) 643--665},
  \href{http://arxiv.org/abs/1711.05594}{{\ttfamily arXiv:1711.05594
  [nucl-ex]}}.

\bibitem{Voloshin:2007pc}
S.~A. Voloshin, A.~M. Poskanzer, A.~Tang, and G.~Wang, ``{Elliptic flow in the
  Gaussian model of eccentricity fluctuations}'',
  \href{http://dx.doi.org/10.1016/j.physletb.2007.11.043}{{\em Phys. Lett. B}
  {\bfseries 659} (2008) 537--541},
  \href{http://arxiv.org/abs/0708.0800}{{\ttfamily arXiv:0708.0800 [nucl-th]}}.

\bibitem{Borghini:2004ra}
N.~Borghini and J.~Y. Ollitrault, ``{Azimuthally sensitive correlations in
  nucleus-nucleus collisions}'',
  \href{http://dx.doi.org/10.1103/PhysRevC.70.064905}{{\em Phys. Rev. C}
  {\bfseries 70} (2004) 064905},
  \href{http://arxiv.org/abs/nucl-th/0407041}{{\ttfamily
  arXiv:nucl-th/0407041}}.

\bibitem{Barlow:2002yb}
R.~Barlow, ``{Systematic errors: Facts and fictions}'', in {\em {Conference on
  Advanced Statistical Techniques in Particle Physics}}, pp.~134--144.
\newblock 7, 2002.
\newblock \href{http://arxiv.org/abs/hep-ex/0207026}{{\ttfamily
  arXiv:hep-ex/0207026}}.

\bibitem{Chen:2020tbl}
W.~Chen, S.~Cao, T.~Luo, L.-G. Pang, and X.-N. Wang, ``{Medium modification of
  $\gamma$-jet fragmentation functions in Pb+Pb collisions at LHC}'',
  \href{http://dx.doi.org/10.1016/j.physletb.2020.135783}{{\em Phys. Lett. B}
  {\bfseries 810} (2020) 135783},
  \href{http://arxiv.org/abs/2005.09678}{{\ttfamily arXiv:2005.09678
  [hep-ph]}}.

\bibitem{Zhao:2021vmu}
W.~Zhao, W.~Ke, W.~Chen, T.~Luo, and X.-N. Wang, ``{From hydro to jet
  quenching, coalescence and hadron cascade: a coupled approach to solving the
  $R_{AA}\otimes v_2$ puzzle}'',
  \href{http://dx.doi.org/10.1103/PhysRevLett.128.022302}{{\em Phys. Rev.
  Lett.} {\bfseries 128} (2022) 022302},
  \href{http://arxiv.org/abs/2103.14657}{{\ttfamily arXiv:2103.14657
  [hep-ph]}}.

\bibitem{ALICE:2013snk}
{\bfseries ALICE} Collaboration, B.~Abelev {\em et~al.}, ``{Long-range angular
  correlations of $\rm \pi$, K and p in p-Pb collisions at $\sqrt{s_{\rm NN}}$
  = 5.02 TeV}'', \href{http://dx.doi.org/10.1016/j.physletb.2013.08.024}{{\em
  Phys. Lett. B} {\bfseries 726} (2013) 164--177},
  \href{http://arxiv.org/abs/1307.3237}{{\ttfamily arXiv:1307.3237 [nucl-ex]}}.

\end{thebibliography}\endgroup

\newpage
\appendix


%
%

\section{The ALICE Collaboration}
\label{app:collab}
\begin{flushleft} 
\small

S.~Acharya\,\orcidlink{0000-0002-9213-5329}\,$^{\rm 125,132}$, 
D.~Adamov\'{a}\,\orcidlink{0000-0002-0504-7428}\,$^{\rm 86}$, 
A.~Adler$^{\rm 69}$, 
G.~Aglieri Rinella\,\orcidlink{0000-0002-9611-3696}\,$^{\rm 32}$, 
M.~Agnello\,\orcidlink{0000-0002-0760-5075}\,$^{\rm 29}$, 
N.~Agrawal\,\orcidlink{0000-0003-0348-9836}\,$^{\rm 50}$, 
Z.~Ahammed\,\orcidlink{0000-0001-5241-7412}\,$^{\rm 132}$, 
S.~Ahmad\,\orcidlink{0000-0003-0497-5705}\,$^{\rm 15}$, 
S.U.~Ahn\,\orcidlink{0000-0001-8847-489X}\,$^{\rm 70}$, 
I.~Ahuja\,\orcidlink{0000-0002-4417-1392}\,$^{\rm 37}$
% ...

\section*{Affiliation Notes}

$^{\rm I}$ Deceased\\
$^{\rm II}$ Also at: Max-Planck-Institut f\"{u}r Physik, Munich, Germany\\
$^{\rm III}$ Also at: Italian National Agency for New Technologies, Energy and Sustainable Economic Development (ENEA), Bologna, Italy\\
$^{\rm IV}$ Also at: Dipartimento DET del Politecnico di Torino, Turin, Italy\\
$^{\rm V}$ Also at: Department of Applied Physics, Aligarh Muslim University, Aligarh, India\\
$^{\rm VI}$ Also at: Institute of Theoretical Physics, University of Wroclaw, Poland\\
$^{\rm VII}$ Also at: An institution covered by a cooperation agreement with CERN\\

\section*{Collaboration Institutes}

$^{1}$ A.I. Alikhanyan National Science Laboratory (Yerevan Physics Institute) Foundation, Yerevan, Armenia\\
$^{2}$ AGH University of Science and Technology, Cracow, Poland\\
$^{3}$ Bogolyubov Institute for Theoretical Physics, National Academy of Sciences of Ukraine, Kiev, Ukraine\\
$^{4}$ Bose Institute, Department of Physics  and Centre for Astroparticle Physics and Space Science (CAPSS), Kolkata, India\\
$^{5}$ California Polytechnic State University, San Luis Obispo, California, United States\\
$^{6}$ Central China Normal University, Wuhan, China\\
$^{7}$ Centro de Aplicaciones Tecnol\'{o}gicas y Desarrollo Nuclear (CEADEN), Havana, Cuba\\
$^{8}$ Centro de Investigaci\'{o}n y de Estudios Avanzados (CINVESTAV), Mexico City and M\'{e}rida, Mexico\\
$^{9}$ Chicago State University, Chicago, Illinois, United States\\
$^{10}$ China Institute of Atomic Energy, Beijing, China\\
$^{11}$ Chungbuk National University, Cheongju, Republic of Korea\\
$^{12}$ Comenius University Bratislava, Faculty of Mathematics, Physics and Informatics, Bratislava, Slovak Republic\\
$^{13}$ COMSATS University Islamabad, Islamabad, Pakistan\\
$^{14}$ Creighton University, Omaha, Nebraska, United States\\
$^{15}$ Department of Physics, Aligarh Muslim University, Aligarh, India\\
$^{16}$ Department of Physics, Pusan National University, Pusan, Republic of Korea\\
$^{17}$ Department of Physics, Sejong University, Seoul, Republic of Korea\\
$^{18}$ Department of Physics, University of California, Berkeley, California, United States\\
$^{19}$ Department of Physics, University of Oslo, Oslo, Norway\\
$^{20}$ Department of Physics and Technology, University of Bergen, Bergen, Norway\\
$^{21}$ Dipartimento di Fisica, Universit\`{a} di Pavia, Pavia, Italy\\
$^{22}$ Dipartimento di Fisica dell'Universit\`{a} and Sezione INFN, Cagliari, Italy\\
$^{23}$ Dipartimento di Fisica dell'Universit\`{a} and Sezione INFN, Trieste, Italy\\
$^{24}$ Dipartimento di Fisica dell'Universit\`{a} and Sezione INFN, Turin, Italy\\
$^{25}$ Dipartimento di Fisica e Astronomia dell'Universit\`{a} and Sezione INFN, Bologna, Italy\\
$^{26}$ Dipartimento di Fisica e Astronomia dell'Universit\`{a} and Sezione INFN, Catania, Italy\\
$^{27}$ Dipartimento di Fisica e Astronomia dell'Universit\`{a} and Sezione INFN, Padova, Italy\\
$^{28}$ Dipartimento di Fisica `E.R.~Caianiello' dell'Universit\`{a} and Gruppo Collegato INFN, Salerno, Italy\\
$^{29}$ Dipartimento DISAT del Politecnico and Sezione INFN, Turin, Italy\\
$^{30}$ Dipartimento di Scienze MIFT, Universit\`{a} di Messina, Messina, Italy\\
$^{31}$ Dipartimento Interateneo di Fisica `M.~Merlin' and Sezione INFN, Bari, Italy\\
$^{32}$ European Organization for Nuclear Research (CERN), Geneva, Switzerland\\
$^{33}$ Faculty of Electrical Engineering, Mechanical Engineering and Naval Architecture, University of Split, Split, Croatia\\
$^{34}$ Faculty of Engineering and Science, Western Norway University of Applied Sciences, Bergen, Norway\\
$^{35}$ Faculty of Nuclear Sciences and Physical Engineering, Czech Technical University in Prague, Prague, Czech Republic\\
$^{36}$ Faculty of Physics, Sofia University, Sofia, Bulgaria\\
$^{37}$ Faculty of Science, P.J.~\v{S}af\'{a}rik University, Ko\v{s}ice, Slovak Republic\\
$^{38}$ Frankfurt Institute for Advanced Studies, Johann Wolfgang Goethe-Universit\"{a}t Frankfurt, Frankfurt, Germany\\
$^{39}$ Fudan University, Shanghai, China\\
$^{40}$ Gangneung-Wonju National University, Gangneung, Republic of Korea\\
$^{41}$ Gauhati University, Department of Physics, Guwahati, India\\
$^{42}$ Helmholtz-Institut f\"{u}r Strahlen- und Kernphysik, Rheinische Friedrich-Wilhelms-Universit\"{a}t Bonn, Bonn, Germany\\
$^{43}$ Helsinki Institute of Physics (HIP), Helsinki, Finland\\
$^{44}$ High Energy Physics Group,  Universidad Aut\'{o}noma de Puebla, Puebla, Mexico\\
$^{45}$ Horia Hulubei National Institute of Physics and Nuclear Engineering, Bucharest, Romania\\
$^{46}$ Indian Institute of Technology Bombay (IIT), Mumbai, India\\
$^{47}$ Indian Institute of Technology Indore, Indore, India\\
$^{48}$ INFN, Laboratori Nazionali di Frascati, Frascati, Italy\\
$^{49}$ INFN, Sezione di Bari, Bari, Italy\\
$^{50}$ INFN, Sezione di Bologna, Bologna, Italy\\
$^{51}$ INFN, Sezione di Cagliari, Cagliari, Italy\\
$^{52}$ INFN, Sezione di Catania, Catania, Italy\\
$^{53}$ INFN, Sezione di Padova, Padova, Italy\\
$^{54}$ INFN, Sezione di Pavia, Pavia, Italy\\
$^{55}$ INFN, Sezione di Torino, Turin, Italy\\
$^{56}$ INFN, Sezione di Trieste, Trieste, Italy\\
$^{57}$ Inha University, Incheon, Republic of Korea\\
$^{58}$ Institute for Gravitational and Subatomic Physics (GRASP), Utrecht University/Nikhef, Utrecht, Netherlands\\
$^{59}$ Institute of Experimental Physics, Slovak Academy of Sciences, Ko\v{s}ice, Slovak Republic\\
$^{60}$ Institute of Physics, Homi Bhabha National Institute, Bhubaneswar, India\\
$^{61}$ Institute of Physics of the Czech Academy of Sciences, Prague, Czech Republic\\
$^{62}$ Institute of Space Science (ISS), Bucharest, Romania\\
$^{63}$ Institut f\"{u}r Kernphysik, Johann Wolfgang Goethe-Universit\"{a}t Frankfurt, Frankfurt, Germany\\
$^{64}$ Instituto de Ciencias Nucleares, Universidad Nacional Aut\'{o}noma de M\'{e}xico, Mexico City, Mexico\\
$^{65}$ Instituto de F\'{i}sica, Universidade Federal do Rio Grande do Sul (UFRGS), Porto Alegre, Brazil\\
$^{66}$ Instituto de F\'{\i}sica, Universidad Nacional Aut\'{o}noma de M\'{e}xico, Mexico City, Mexico\\
$^{67}$ iThemba LABS, National Research Foundation, Somerset West, South Africa\\
$^{68}$ Jeonbuk National University, Jeonju, Republic of Korea\\
$^{69}$ Johann-Wolfgang-Goethe Universit\"{a}t Frankfurt Institut f\"{u}r Informatik, Fachbereich Informatik und Mathematik, Frankfurt, Germany\\
$^{70}$ Korea Institute of Science and Technology Information, Daejeon, Republic of Korea\\
$^{71}$ KTO Karatay University, Konya, Turkey\\
$^{72}$ Laboratoire de Physique des 2 Infinis, Ir\`{e}ne Joliot-Curie, Orsay, France\\
$^{73}$ Laboratoire de Physique Subatomique et de Cosmologie, Universit\'{e} Grenoble-Alpes, CNRS-IN2P3, Grenoble, France\\
$^{74}$ Lawrence Berkeley National Laboratory, Berkeley, California, United States\\
$^{75}$ Lund University Department of Physics, Division of Particle Physics, Lund, Sweden\\
$^{76}$ Nagasaki Institute of Applied Science, Nagasaki, Japan\\
$^{77}$ Nara Women{'}s University (NWU), Nara, Japan\\
$^{78}$ National and Kapodistrian University of Athens, School of Science, Department of Physics , Athens, Greece\\
$^{79}$ National Centre for Nuclear Research, Warsaw, Poland\\
$^{80}$ National Institute of Science Education and Research, Homi Bhabha National Institute, Jatni, India\\
$^{81}$ National Nuclear Research Center, Baku, Azerbaijan\\
$^{82}$ National Research and Innovation Agency - BRIN, Jakarta, Indonesia\\
$^{83}$ Niels Bohr Institute, University of Copenhagen, Copenhagen, Denmark\\
$^{84}$ Nikhef, National institute for subatomic physics, Amsterdam, Netherlands\\
$^{85}$ Nuclear Physics Group, STFC Daresbury Laboratory, Daresbury, United Kingdom\\
$^{86}$ Nuclear Physics Institute of the Czech Academy of Sciences, Husinec-\v{R}e\v{z}, Czech Republic\\
$^{87}$ Oak Ridge National Laboratory, Oak Ridge, Tennessee, United States\\
$^{88}$ Ohio State University, Columbus, Ohio, United States\\
$^{89}$ Physics department, Faculty of science, University of Zagreb, Zagreb, Croatia\\
$^{90}$ Physics Department, Panjab University, Chandigarh, India\\
$^{91}$ Physics Department, University of Jammu, Jammu, India\\
$^{92}$ Physics Department, University of Rajasthan, Jaipur, India\\
$^{93}$ Physics Program and International Institute for Sustainability with Knotted Chiral Meta Matter (SKCM2), Hiroshima University, Hiroshima, Japan\\
$^{94}$ Physikalisches Institut, Eberhard-Karls-Universit\"{a}t T\"{u}bingen, T\"{u}bingen, Germany\\
$^{95}$ Physikalisches Institut, Ruprecht-Karls-Universit\"{a}t Heidelberg, Heidelberg, Germany\\
$^{96}$ Physik Department, Technische Universit\"{a}t M\"{u}nchen, Munich, Germany\\
$^{97}$ Politecnico di Bari and Sezione INFN, Bari, Italy\\
$^{98}$ Research Division and ExtreMe Matter Institute EMMI, GSI Helmholtzzentrum f\"ur Schwerionenforschung GmbH, Darmstadt, Germany\\
$^{99}$ Saga University, Saga, Japan\\
$^{100}$ Saha Institute of Nuclear Physics, Homi Bhabha National Institute, Kolkata, India\\
$^{101}$ School of Physics and Astronomy, University of Birmingham, Birmingham, United Kingdom\\
$^{102}$ Secci\'{o}n F\'{\i}sica, Departamento de Ciencias, Pontificia Universidad Cat\'{o}lica del Per\'{u}, Lima, Peru\\
$^{103}$ Stefan Meyer Institut f\"{u}r Subatomare Physik (SMI), Vienna, Austria\\
$^{104}$ SUBATECH, IMT Atlantique, Nantes Universit\'{e}, CNRS-IN2P3, Nantes, France\\
$^{105}$ Suranaree University of Technology, Nakhon Ratchasima, Thailand\\
$^{106}$ Technical University of Ko\v{s}ice, Ko\v{s}ice, Slovak Republic\\
$^{107}$ The Henryk Niewodniczanski Institute of Nuclear Physics, Polish Academy of Sciences, Cracow, Poland\\
$^{108}$ The University of Texas at Austin, Austin, Texas, United States\\
$^{109}$ Universidad Aut\'{o}noma de Sinaloa, Culiac\'{a}n, Mexico\\
$^{110}$ Universidade de S\~{a}o Paulo (USP), S\~{a}o Paulo, Brazil\\
$^{111}$ Universidade Estadual de Campinas (UNICAMP), Campinas, Brazil\\
$^{112}$ Universidade Federal do ABC, Santo Andre, Brazil\\
$^{113}$ University of Cape Town, Cape Town, South Africa\\
$^{114}$ University of Houston, Houston, Texas, United States\\
$^{115}$ University of Jyv\"{a}skyl\"{a}, Jyv\"{a}skyl\"{a}, Finland\\
$^{116}$ University of Kansas, Lawrence, Kansas, United States\\
$^{117}$ University of Liverpool, Liverpool, United Kingdom\\
$^{118}$ University of Science and Technology of China, Hefei, China\\
$^{119}$ University of South-Eastern Norway, Kongsberg, Norway\\
$^{120}$ University of Tennessee, Knoxville, Tennessee, United States\\
$^{121}$ University of the Witwatersrand, Johannesburg, South Africa\\
$^{122}$ University of Tokyo, Tokyo, Japan\\
$^{123}$ University of Tsukuba, Tsukuba, Japan\\
$^{124}$ University Politehnica of Bucharest, Bucharest, Romania\\
$^{125}$ Universit\'{e} Clermont Auvergne, CNRS/IN2P3, LPC, Clermont-Ferrand, France\\
$^{126}$ Universit\'{e} de Lyon, CNRS/IN2P3, Institut de Physique des 2 Infinis de Lyon, Lyon, France\\
$^{127}$ Universit\'{e} de Strasbourg, CNRS, IPHC UMR 7178, F-67000 Strasbourg, France, Strasbourg, France\\
$^{128}$ Universit\'{e} Paris-Saclay Centre d'Etudes de Saclay (CEA), IRFU, D\'{e}partment de Physique Nucl\'{e}aire (DPhN), Saclay, France\\
$^{129}$ Universit\`{a} degli Studi di Foggia, Foggia, Italy\\
$^{130}$ Universit\`{a} del Piemonte Orientale, Vercelli, Italy\\
$^{131}$ Universit\`{a} di Brescia, Brescia, Italy\\
$^{132}$ Variable Energy Cyclotron Centre, Homi Bhabha National Institute, Kolkata, India\\
$^{133}$ Warsaw University of Technology, Warsaw, Poland\\
$^{134}$ Wayne State University, Detroit, Michigan, United States\\
$^{135}$ Westf\"{a}lische Wilhelms-Universit\"{a}t M\"{u}nster, Institut f\"{u}r Kernphysik, M\"{u}nster, Germany\\
$^{136}$ Wigner Research Centre for Physics, Budapest, Hungary\\
$^{137}$ Yale University, New Haven, Connecticut, United States\\
$^{138}$ Yonsei University, Seoul, Republic of Korea\\
$^{139}$  Zentrum  f\"{u}r Technologie und Transfer (ZTT), Worms, Germany\\
$^{140}$ Affiliated with an institute covered by a cooperation agreement with CERN\\
$^{141}$ Affiliated with an international laboratory covered by a cooperation agreement with CERN.\\

\end{flushleft} 
  
\end{document}